\documentclass[12pt,preprint]{emulateapj}
\usepackage{epsfig}
\usepackage{amsmath}
\usepackage{natbib}

\usepackage{lineno}

\def \teff {$T_{\rm eff}$}
\def \logteff {$\rm log(T_{\rm eff})$}
\def \logg {$\log (g)$}

\def \rev {}

\begin{document}


\title{Helium-Abundance and Other Composition Effects on the Properties of Stellar Surface Convection in Solar-like Main-sequence Stars}
\author{Joel D. Tanner, Sarbani Basu \& Pierre Demarque}
\affil{Department of Astronomy, Yale University, P.O. BOX 208101, New Haven, CT 06520-8101}
{\email{joel.tanner@yale.edu}}


\begin{abstract}
We investigate the effect of helium abundance and $\alpha$-element enhancement on the properties of convection in envelopes of solar-like main-sequence stars stars using a grid of 3D radiation hydrodynamic simulations.  Helium abundance increases the mean molecular weight of the gas, and alters opacity by displacing hydrogen.  Since the scale of the effect of helium may depend on the metallicity, the grid consists of simulations with three helium abundances ($Y=0.1, 0.2, 0.3$), each with two metallicities ($Z=0.001, 0.020)$.   We find that changing the helium mass fraction generally affects structure and convective dynamics in a way opposite to that of metallicity.  Furthermore, the effect is considerably smaller than that of metallicity.  The signature of helium differs from that of metallicity in the manner in which the photospheric velocity distribution is affected.  \rev{We also find that helium abundance and surface gravity behave largely in similar ways, but differ in the way they affect the mean molecular weight}.  A simple model for spectral line formation suggests that the bisectors and absolute Doppler shifts of spectral lines depends on the helium abundance. We look at the effect of $\alpha$-element enhancement and find that it has a considerably smaller effect on the convective dynamics in the SAL compared to that of helium abundance.

\end{abstract}

\section{Introduction} \label{sec:introduction}

Understanding stellar convection and accurately representing it in stellar models remains a formidable challenge.  One of the commonly used treatments for convection is the mixing length theory \citep[MLT;][]{1958ZA.....46..108B},\ which represents convection with a characteristic length scale.  The mixing length parameter (which is the ratio of the mixing length to the pressure scale height, $\alpha = l / H_P$) sets the specific entropy of the convection zone, which in turn adjusts the stellar radius.  This provides a means of determining the mixing length parameter for the Sun, since we have precise constraints on the solar radius.  This solar-calibrated value for the mixing length parameter is usually used to model all other stars, but there is evidence suggesting that this is not the case in reality.

Even when properly calibrated to the stellar radius, MLT-like treatments of stellar convection fail to accurately represent inefficient convection.  This failure severely limits the accuracy of the convective envelope boundaries within stellar models.  One such boundary is the superadiabatic layer (SAL), which is a region near the surface of stars with convective envelopes.  This layer spans several scale heights and encompasses the transition from efficient convective energy transport to radiative.

Radiation hydrodynamic (RHD) simulations of the SAL have proven to be a useful tool for studying realistic stellar convection.  Simulations self-consistently couple the convective envelope to the radiative atmosphere, while including important physical processes such as turbulent pressure.  Established by \citet{1982A&A...107....1N,1985SoPh..100..209N} as a feasible technique for studying realistic stellar convection in the SAL, RHD simulations have been carried by a number of groups.  \citep[e.g.][etc.]{1989ApJ...336.1022C, 1989ApJ...342L..95S, 1998ApJ...499..914S, 2000SoPh..192...91S, 1991ApJ...370..282C,  1995ApJ...442..422K, 2003MNRAS.340..923R, 2004MNRAS.347.1208R, 2005MNRAS.362.1031R, 2007ASPC..362..306J}. Efforts to methodically simulate convection in the $\log(g)$-$\log(T_{\rm eff})$ plane \citep[e.g.][etc.]{1995LIACo..32..213L, 1998IAUS..185..115L, 1999A&A...346..111L, 1999ASPC..173..225F, 2009MmSAI..80..711L, 2011JPhCS.328a2003C} reveal a significant and systematic variation in convective properties over a broad range in stellar properties.

In addition to stellar properties, the convective dynamics near the stellar surface are affected by the chemical composition of the convective envelope.  The recent observational study by \citet{2012ApJ...755L..12B} and theoretical work by \citet{2013ApJ...767...78T}  both suggest that the stellar surface convection depends on the chemical composition of the convective envelope.  Similar results are obtained by \citet{2013arXiv1302.2621M}. \citet{2009CoAst.160...30K} examined helium abundance in 2D simulations, and found that the helium-free case was characterized by larger velocities and flow structures.  The abundance of helium and its effect on stellar convection, however, has not yet been examined systematically.  In this study we build upon our previous work by adding helium abundance to the parameter space of our simulation grid, and isolate its effect on convection by fixing the other dimensions of parameter space.

It is difficult to measure spectroscopically the helium mass fraction present in stars that are cool enough to have convective envelopes. Asteroseismology permits constraints on the helium abundance through measurements of low degree acoustic modes \citep{2004ESASP.559..313B,2007MNRAS.375..861H}, however helium abundance in stars remains largely unconstrained.  The common practice is to use a chemical evolution formula, such as that of \citet{2008ApJS..178...89D}. Quite separately from chemical evolution, the helium content may be enriched or depleted in stellar envelopes through processes such as rotationally induced mixing and meridional circulation \citep[e.g.][]{1979ApJ...229..624S, 1992A&A...265..115Z, 1997ApJ...474L..23S}. For example, the multiple stellar populations present in $\omega$ Centauri can be modeled if rather extreme helium abundances are used.  A recent study by \citet{eschpaper} finds that helium mass fractions in excess of $Y=0.40$ yield the best models for certain EHB stars.  Conversely, helium could be depleted in the outer layers of stars as a result of diffusion from the gravitational settling \citep[][etc]{1992ApJ...394..515C}. 

In Sections \ref{sec:code} and \ref{sec:grid} describe the 3D RHD code and the grid of simulations used in this work.  Section \ref{sec:microphysics} outlines the changes to the microphysics induced by altering the hydrogen mass fraction, and the expected effect that these changes will have on convective dynamics.  We then examine in detail the consequences of hydrogen abundance on various properties extracted from the simulations in Sections \ref{sec:salstructure} through \ref{sec:spectral}.  \rev{We briefly discuss the significance of the $\alpha$-enhancement in the context of globular cluster environments in Section \ref{sec:alpha}.  Finally, in Section \ref{sec:gravity}, we examine the extent to which {\logg} mimics helium abundance in 3D simulations, as it is known to do in 1D model atmospheres. }

\section{The Radiation Hydrodynamics Code} \label{sec:code}

Our code simulates convection by solving the compressible Navier-Stokes equations with radiative transfer, and has been described in detail in \cite{2012ApJ...759..120T}.  The code is originally based on that of \citet{1989ApJ...336.1022C} and \citet{1998ApJ...496L.121K} but has updated physics and numerical schemes.

The simulation code uses microphysics that are fully consistent with that of the stellar evolution code YREC \citep{2008Ap&SS.316...31D}.  The equation of state and opacity are taken from the OPAL  tables \citep{2002ApJ...576.1064R} and the \citet{2005ApJ...623..585F} opacity tables at low temperatures using the heavy element mixture of \citet{1998SSRv...85..161G}.  For the simulations in this work, radiative transfer was computed using the 3D Eddington approximation \citep{1966PASJ...18...85U} in the optically thin layers, and the diffusion approximation in the deeper optically thick layers, however, as \citet{2012ApJ...759..120T} showed, using a different radiative transfer scheme (rays with long-characteristics) does not make a substantial difference in the SAL below the photosphere.

The computational domain is a Cartesian box with periodic boundary conditions on the vertical walls, and the top and bottom surfaces are closed.  Spatial curvature and the radial variation of gravity are both negligible.  The domain spans the SAL so the bottom surface is well below the superadiabatic peak, and the top surface is in the optically thin radiative layers of the atmosphere.  Thus, the space between the two surfaces encompasses the transition from fully convective to fully radiative energy transport. 

\rev{Regardless of implementation, boundary conditions will introduce unwanted artifical effects to the dynamics and structure of the simulation.  \citet{2007MNRAS.374..305K} and \citet{2013arXiv1305.0743G} suggest that the domain of influence of the closed boundary can extend one or two pressure scale heights from the boundary surface.  Thus, to minimize the effect of the boundary surfaces on our analysis, we trim one scale height from the top and bottom of the simulation domain.   The complete simulation domain extends one pressure scale height above what is shown in all figures presented in subsequent sections, and the lower boundary layer is below the plotted range.}

\section{The Grid of Simulations} \label{sec:grid}

The strategy for exploring the effect of helium on convection closely mirrors the approach described in \citet{2013ApJ...767...78T}, which dealt only with changes to the mass fraction of metal elements. In this study, we construct a grid of simulations that covers a range in helium mass fractions.  Because the effect of helium may differ between metal-rich and metal-poor stars, we compute our simulations with different hydrogen mass fractions ($Y=0.1$, $0.2$, $0.3$) for two different metallicities ($Z=0.001$, $Z=0.020$). The grid of simulations comprises six sets, each with a different chemical composition.  The grid is divided into two groups (each with three sets), corresponding to low and high metallicity.  The sets within a group have fixed metallicity, but different helium abundances.  The range in helium and metallicity is quite large, covering a factor of three in $Y$ and a factor of $20$ in $Z$.  While large, the range in helium mass fraction is less extreme than what has been used to model some stellar populations, including that of $\omega$ Centauri, which \citet{eschpaper} recently modeled with helium enrichments of up to $Y=0.40$.  

The grid of 24 simulations is summarized in Table \ref{tab:grid}.  The two metallicity groups are denoted with ‘a’ and ‘b’ for the low- and high-metallicity, respectively.  Surface gravity is held fixed at $\log(g)=4.30$ for the entire grid.  The effective temperature for each simulation is an output property, and cannot be precisely controlled.  As such, simulations typically cannot be directly compared, but each set overlaps in the effective temperature range, allowing for comparisons as a function of radiative flux.  

\begin{table*}[t]
  \centering
  \caption{Properties of the simulations in the grid.  All simulations have the same surface gravity ($\log(g) = 4.30$).  The grid is divided into two parts (low and high $Z$), with each part containing three sets at with different $Y$.}
  \label{tab:grid}
  \begin{center}
    \leavevmode
    \begin{tabular}{cccccccc} \hline \hline              
  $ID$ & Z & Y & $\log T_{\rm eff}$ & $\delta x \times \delta z$ (km) & $N_x \times N_z$ & $\Delta x$ (Mm) & $\Delta z$ (Mm)  \\ \hline 
s01a & 0.001 & 0.100 & 3.735 & $50.23 \times 16.07$ & $95 \times 205$ & 4.77 & 3.29  \\
s02a & 0.001 & 0.100 & 3.751 & $53.56 \times 17.13$ & $95 \times 205$ & 5.09 & 3.51  \\
s03a & 0.001 & 0.100 & 3.765 & $57.88 \times 18.52$ & $95 \times 210$ & 5.50 & 3.89  \\
s04a & 0.001 & 0.100 & 3.780 & $63.33 \times 20.26$ & $95 \times 210$ & 6.02 & 4.25  \\
s05a & 0.001 & 0.200 & 3.752 & $50.23 \times 16.07$ & $95 \times 205$ & 4.77 & 3.29  \\
s06a & 0.001 & 0.200 & 3.765 & $53.56 \times 17.13$ & $95 \times 205$ & 5.09 & 3.51  \\
s07a & 0.001 & 0.200 & 3.778 & $57.88 \times 18.52$ & $95 \times 210$ & 5.50 & 3.89  \\
s08a & 0.001 & 0.200 & 3.790 & $63.33 \times 20.26$ & $95 \times 210$ & 6.02 & 4.25  \\
s09a & 0.001 & 0.300 & 3.766 & $50.23 \times 16.07$ & $95 \times 205$ & 4.77 & 3.29  \\
s10a & 0.001 & 0.300 & 3.778 & $53.56 \times 17.13$ & $95 \times 205$ & 5.09 & 3.51  \\
s11a & 0.001 & 0.300 & 3.789 & $57.88 \times 18.52$ & $95 \times 210$ & 5.50 & 3.89  \\
s12a & 0.001 & 0.300 & 3.800 & $63.33 \times 20.26$ & $95 \times 210$ & 6.02 & 4.25  \\
s01b & 0.020 & 0.100 & 3.676 & $50.23 \times 16.07$ & $95 \times 205$ & 4.77 & 3.29  \\
s02b & 0.020 & 0.100 & 3.706 & $53.56 \times 17.13$ & $95 \times 205$ & 5.09 & 3.51  \\
s03b & 0.020 & 0.100 & 3.726 & $57.88 \times 18.52$ & $95 \times 210$ & 5.50 & 3.89  \\
s04b & 0.020 & 0.100 & 3.751 & $63.33 \times 20.26$ & $95 \times 210$ & 6.02 & 4.25  \\
s05b & 0.020 & 0.200 & 3.699 & $50.23 \times 16.07$ & $95 \times 205$ & 4.77 & 3.29  \\
s06b & 0.020 & 0.200 & 3.723 & $53.56 \times 17.13$ & $95 \times 205$ & 5.09 & 3.51  \\
s07b & 0.020 & 0.200 & 3.744 & $57.88 \times 18.52$ & $95 \times 210$ & 5.50 & 3.89  \\
s08b & 0.020 & 0.200 & 3.764 & $63.33 \times 20.26$ & $95 \times 210$ & 6.02 & 4.25  \\
s09b & 0.020 & 0.300 & 3.722 & $50.23 \times 16.07$ & $95 \times 205$ & 4.77 & 3.29  \\
s10b & 0.020 & 0.300 & 3.740 & $53.56 \times 17.13$ & $95 \times 205$ & 5.09 & 3.51  \\
s11b & 0.020 & 0.300 & 3.760 & $57.88 \times 18.52$ & $95 \times 210$ & 5.50 & 3.89  \\
s12b & 0.020 & 0.300 & 3.777 & $63.33 \times 20.26$ & $95 \times 210$ & 6.02 & 4.25  \\ \hline
    \end{tabular}
  \end{center}
\end{table*}

Some simulations, however, serendipitously fall very near to each other in effective temperature, which permits direct comparison of their properties as a function of depth without resorting to interpolation.  One example of this in our grid is the group comprised of simulations s4a, s10a, and s12b.  These three simulations reveal the effect of changing the helium mass fraction (comparing s4a and s10a) and the effect of metallicity (comparing s10a and s12b).  These three simulations are showcased throughout the following sections when we examine various properties.  The differences in chemical composition between these three simulations represents much of the range that might be expected in stars near the main sequence. 

\section{Why We Expect Chemical Composition to Change Convective Properties} \label{sec:microphysics}

Altering the chemical composition of the convective envelope introduces change to the microphysics that have a feedback on the convective dynamics.  Changes to the microphysics appear in the opacity and the equation of state, and the significance of these changes derives from the original motivation for 3D simulations, which is to determine the stratification through the SAL.  As mentioned in Section \ref{sec:introduction}, this is usually accomplished through MLT and the selection of the (unconstrained) mixing length parameter.  A range of adiabatic stratifications can be made to be compatible with a particular atmospheric structure by selecting different MLT parameters.  Aside from having an unconstrained free parameter, MLT also fails to include some relevant physics, such as turbulent pressure.

The atmospheric helium mass fraction potentially affects the convective gas dynamics by altering the  opacity, and the equation of state.  While the continuum opacity is not as strongly dependent on hydrogen as metallicity, the effect is nonetheless apparent. The convective motions are driven by the requirement to carry the stellar energy flux through the convective envelope, and the excess weight provided by helium can alter the flow properties.  Enriching or depleting the helium abundance, which alters both the mean molecular weight and the opacity,  is therefore expected to change the convective length scales and the turbulent pressure contribution to hydrostatic equilibrium.

Performing 3D RHD simulations provides a way of self-consistently determining which adiabatic stratification is compatible with a given atmosphere (with a specified surface gravity, effective temperature and composition).  From this perspective, helium composition is important because it changes which adiabatic structure is compatible with a given atmosphere.  Further details are provided in section \ref{sec:meanstratification}, which examines the mean stratification of simulations with varied composition.

\subsection{Opacity}  \label{sec:opacity}

The primary source of continuum opacity for these stars is from the negative hydrogen ion.  This source of opacity depends on the hydrogen mass fraction ($X$), which forms the negative ion, as well as the metal element mass fraction ($Z$), which provide the free electrons as a result of their low ionization energy \citep[e.g.][]{graybook}.  The helium mass fraction ($Y$) will have an effect on the continuum opacity primarily by displacing hydrogen.  While the effect is not as significant as that induced by changes to metallicity, it is large enough to change the structure of the superadiabatic layer.

\begin{figure}[h]
\epsscale{1.20}
\plotone{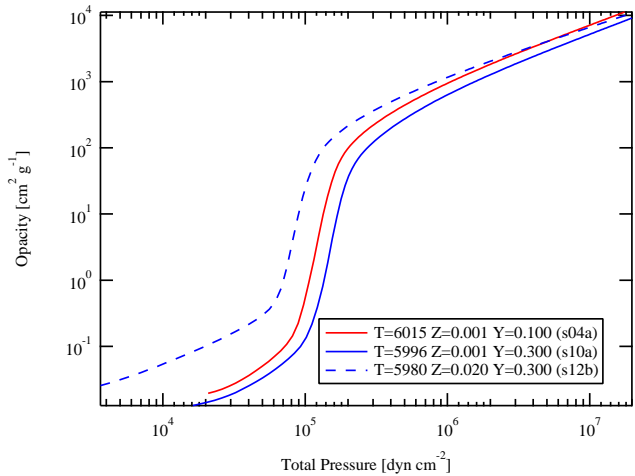}
\caption{Opacity as a function of total pressure for three simulations from the grid.  The magnitude of the shift in opacity as a result of metallicity is greater than that of helium, but is significant in both cases.  Line color and style indicate helium and metal abundance, respectively.}
\label{fig:opacity}
\end{figure}

Fig. \ref{fig:opacity} shows how the Rosseland mean opacity changes when the helium and metal element mass fractions are adjusted.  The opacity is presented as a function of pressure, having been averaged spatially in the horizontal dimensions of the simulation domain.  Opacity is clearly most sensitive to the metallicity, but the effect of helium is non-negligible.   An increase in helium (decrease in hydrogen) changes the opacity in the opposite way as an increase in metallicity.

The location of the SAL is marked by a precipitous drop in the opacity.  The SAL of the high-metallicity simulation is pushed higher into the atmosphere (to lower pressure and density) as a result of the increased opacity.  The effect of decreasing helium is similar but more subdued.

\subsection{Equation of State} \label{sec:eos}

Although the helium abundance does not contribute to the opacity as strongly as metallicity, it has a more significant effect on the equation of state.  In particular, the mean molecular weight is increased substantially by displacing hydrogen with the helium.  Since helium doesn't contribute strongly to opacity, to rough approximation this excess material behaves like `dead weight'.

We compare the molecular weight of gas in the simulations by assuming an ideal gas, where the ratio of the gas constant to the mean molecular weight is  
\begin{equation}
\frac{\mathcal{R}}{\mu} = \frac{P}{\rho T}.
\end{equation}

In this representation, all non-ideal effects in the equation of state are implicitly included in the $\mathcal{R}/\mu$ term.   \rev{We emphasize that the simulations are computed using the OPAL equation of state (which includes non-idealized effects), and that this ideal representation is only used for convenience when calculating the mean molecular weight to analyze the simulation data.}

\begin{figure}[h]
\epsscale{1.20}
\plotone{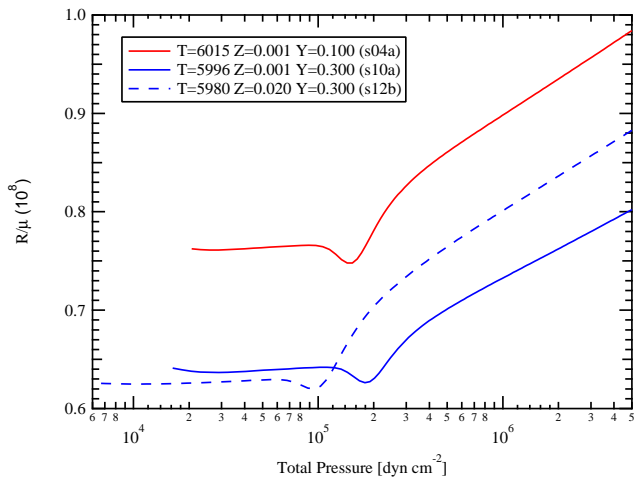}
\caption{Mean molecular weight as a function of total pressure for three simulations from the grid.  Metallicity primarily changes the mean molecular weight below the SAL, while helium abundance shifts it throughout the simulation domain.}
\label{fig:mmwt}
\end{figure}

Fig. \ref{fig:mmwt} shows that changes to metallicity primarily affect the mean molecular weight in the deeper layers.  Changes to the helium mass fraction introduce a shift in the mean molecular weight across the entire range of the simulation domain, including the peak of the SAL and the optically thin atmospheric layers.  This rather large shift in the molecular weight results in a corresponding adjustment to the convective velocity field, resulting in differences in mean velocity and RMS velocity (see Section \ref{sec:dynamics} for a detailed comparison). 

Also included in the equation of state is the effect of ionization.  Particularly relevant to this study is the hydrogen ionization zone, and the first ionization zone of helium.  While the metals can begin to be ionized at low temperatures contributing to the opacity in the atmosphere, the hydrogen ionization zone extends below the peak of the SAL.  Deeper still are the first and second helium ionization regions, which peak near $20000$K and $60000$K, respectively.  The simulation domain extends deep enough to partially capture the first ionization zone of He, which is situated in the near-adiabatic region at the bottom of the box. The second ionization zone is well outside the simulation domain, and so its effect on convection cannot be included.  Because of practical computational limitations, these simulations cannot encompass the entire second ionization zone.

\begin{figure}
\epsscale{1.2}
\begin{minipage}[b]{1.0\linewidth}
\plotone{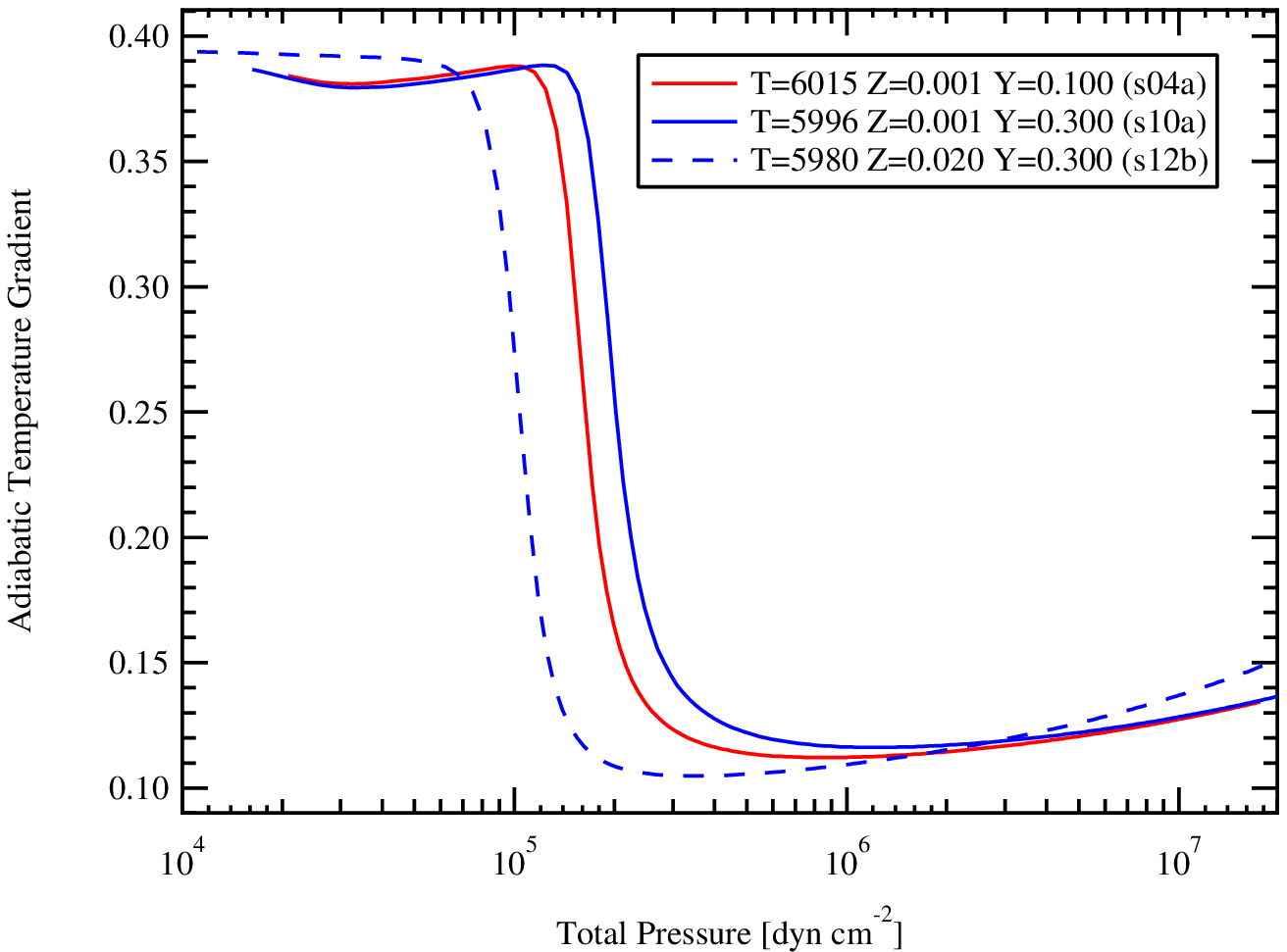}
\epsscale{1.2}
\plotone{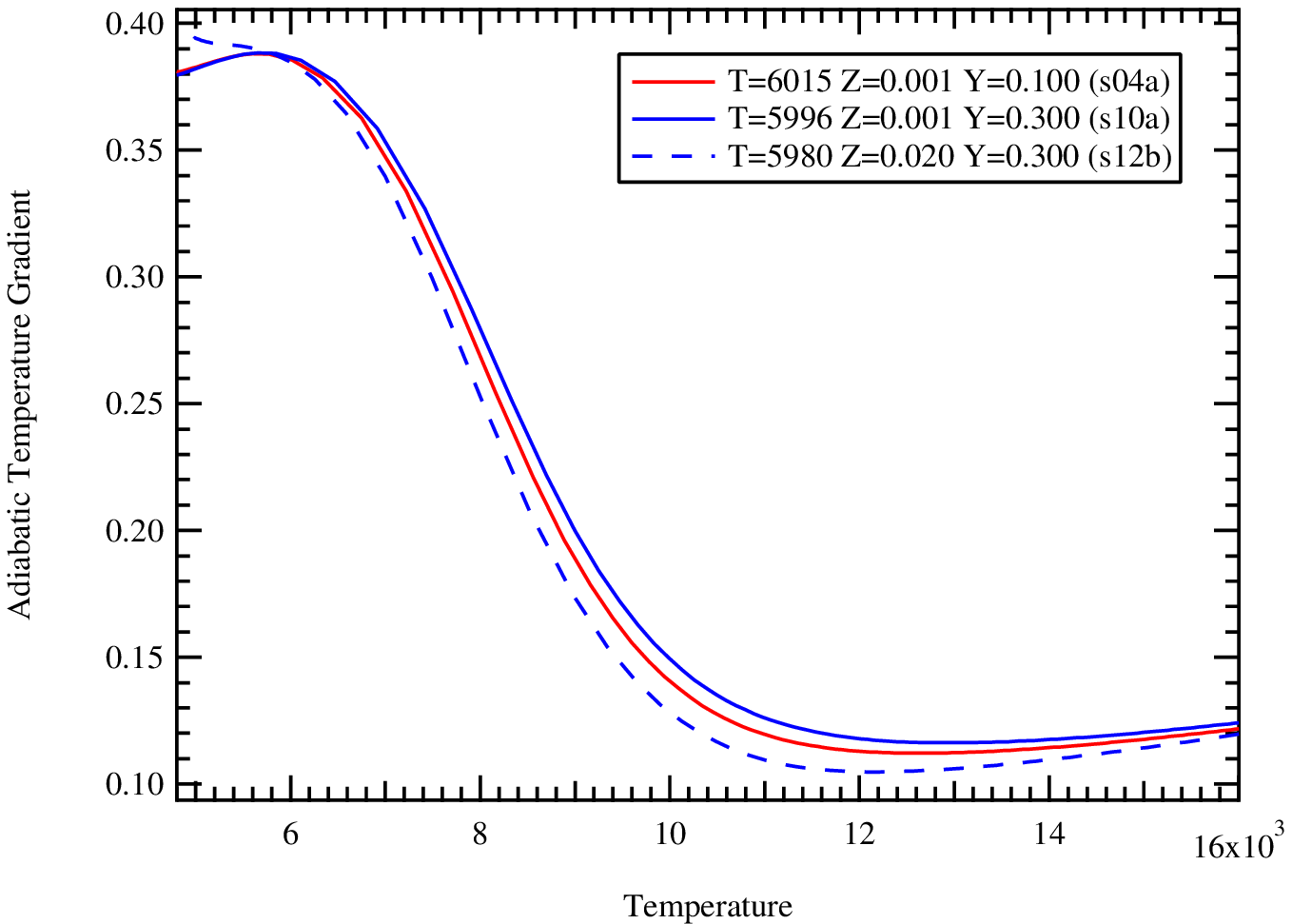}
\end{minipage}
\caption{Adiabatic temperature gradient as a function of total pressure (top) and temperature (bottom) for three simulations from the grid.  Both metallicity and helium abundance shift the location of the broad dip, which corresponds to the hydrogen ionization zone.  Metallicity has a comparatively larger effect than helium, and also changes the adiabatic temperature gradient in the atmosphere above the SAL.}
\label{fig:gradad}
\end{figure}

The adiabatic temperature gradient, $\nabla_{\rm ad}$, is sensitive to the helium and metal mass fraction, as shown in Fig. \ref{fig:gradad}. The broad dip in the gradient through most of the simulation domain is the hydrogen ionization zone.  The effect of the first helium ionization zone is much smaller than that of hydrogen ionization, and not visible in the figure. The adiabatic temperature gradient increases with the fraction of ionized helium because the helium has displaced hydrogen.  The second ionization region sets the slope of the adiabatic gradient at higher temperatures.  As the mass fraction of helium increases, so does its effect on the adiabatic gradient.

Metallicity has comparatively a greater effect on the adiabatic gradient than displacing hydrogen with helium.  The hydrogen ionization zone is affected by the electron pressure, which is altered by changing the mass fraction of metal. Figure \ref{fig:gradad} shows that this effect is stronger than that of changing the helium (or hydrogen) mass fraction.  Changes to metallicity has an effect on the adiabatic temperature gradient through the entire simulation domain, while changing the helium abundance does not have a significant effect in the atmospheric layers above the SAL. Because helium and metallicity affect different domains, comparing the adiabatic gradient above and below the SAL could provide a way of distinguishing between the two effects.

\section{Structure of the Superadiabatic Layer} \label{sec:salstructure}

\subsection{Mean Stratification} \label{sec:meanstratification}

The stratification from the 3D simulations is the result of self-consistently coupling the radiative transfer in optically thin layers with the convectively efficient adiabatic region below.  The SAL in between includes the turbulent pressure support from inefficient convection.

Examining the mean stratification of the simulation shows how the fully convective deep stratification can be self-consistently coupled to the SAL and atmosphere.  The construction of 1D models typically utilizes an approximation similar to the mixing length theory to set the structure of the SAL.  When using MLT-like treatments, a wide range of adiabatic structures is compatible with a given atmosphere, depending on the value of the mixing length parameter.  Simulations provide realistic stratifications through the region of transition from convective to radiative energy transport, and provide insight into what improvements might be made to existing treatments for convection in stellar models.

\begin{figure}[h]
\epsscale{1.2}
\plotone{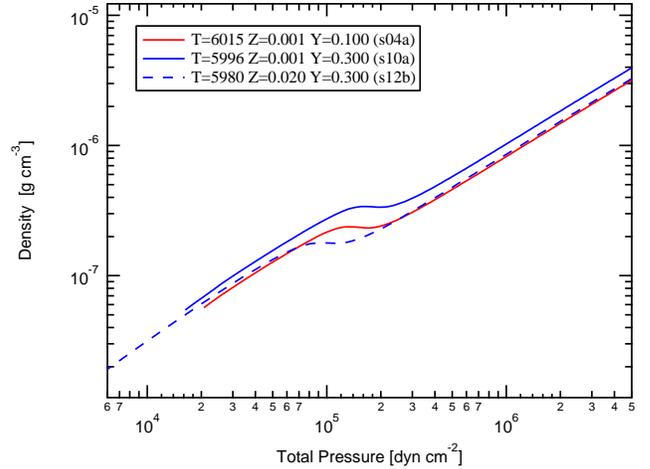}
\caption{Mean stratifications from three simulations in the grid.  The surface gravity and effective temperature are essentially the same, so any differences in structure are attributed to composition.  One pair of simulations differ only in $Z$, while the other pair differ only in $Y$.}
\label{fig:rhovsp}
\end{figure}

For a given {\logteff} and {\logg}, the structure of the atmosphere is determined by the chemical composition.  Increasing the opacity results in a lower density through the SAL so that the energy flux is maintained.  As described in \citet{2013ApJ...767...78T} and presented in Figure \ref{fig:rhovsp}, a lower density in the SAL requires a lower density in the adiabatic structure below.  Note that this behaviour is not present in MLT models, where the density of the adiabatic structure is determined by the value of the mixing length parameter.

In Section \ref{sec:microphysics} we outlined the changes to the microphysics as a result of the helium abundance.  The combination of changing the mean molecular weight, opacity, and equation of state result in a different pairing of an adiabatic structure with an atmosphere.  Fig. \ref{fig:rhovsp} compares density as a function of pressure in three simulations that have the same {\logteff} and {\logg}.  Lowering the metallicity significantly reduces the opacity, resulting in a corresponding rise in the density across the entire simulation domain.  

Generally, adjusting the helium abundance has a similar effect as that of metallicity, but in the opposite direction, which is consistent with the change in opacity.  For example, comparing in Fig. \ref{fig:rhovsp}, from comparing the high-$Z$ high-$Y$ structure with that of the low-$Z$ low-$Y$ composition it is clear that the adiabatic and atmospheric structures can remain essentially unchanged over a range in metallicity, provided that a corresponding adjustment is made to the helium abundance.  Note, however, that the structures near the SAL maximum (denoted by a change in slope in density as a function of pressure) are distinct, despite having essentially the same adiabat and atmosphere.  Put another way, for a given atmospheric and adiabatic structure, the helium abundance is instrumental in determining the structure of the SAL between optically thin atmosphere and the deeper adiabatic layers.

\subsection{Superadiabatic Excess ($\nabla - \nabla_{\rm ad}$)} \label{sec:pturb}

Comparisons of the mean structure in Section \ref{sec:meanstratification} revealed that the structure of SAL depends on both metallicity and helium abundance.  For a given {\logteff} and {\logg}, the structure through the SAL is determined by the transition from convective to radiative energy transport.  As evidenced by the different structures presented in Figure \ref{fig:rhovsp}, the rate and nature of the transition clearly depends on composition. The difference between the actual temperature gradient and adiabatic temperature gradient, or superadiabaitic excess, is a measure of convective efficiency.  The location and extent of this excess temperature gradient reveals the location of the transition from convective to radiative energy transport.  

\rev{We compute the adiabatic temperature gradient $\nabla_{\rm ad}$ from the OPAL equation of state tables using temperature and gas pressure from the simulation.}  The adiabatic temperature gradient is defined as:
\begin{equation}
\nabla_{\rm ad} = \left( \frac{d \ln T}{d \ln P_{\rm gas}} \right)_{\rm ad}.
\end{equation}

\rev{The actual temperature gradient is computed directly from the simulation with the total pressure ($P_{\rm tot} = P_{\rm gas} + P_{\rm turb}$), which includes turbulent support in the pressure term defined in Equation \ref{eqn:pturb}.}  The dimensionless temperature gradient is defined as:
\begin{equation}
\nabla = \left( \frac{d \ln T}{d \ln P_{\rm tot}} \right).
\end{equation}

The location of the SAL as a function of pressure changes in response to the opacity.  An increase in $Y$ pushes the SAL in the opposite manner as an increase in $Z$.   

\begin{figure}
\epsscale{1.2}
\begin{minipage}[b]{1.0\linewidth}
\plotone{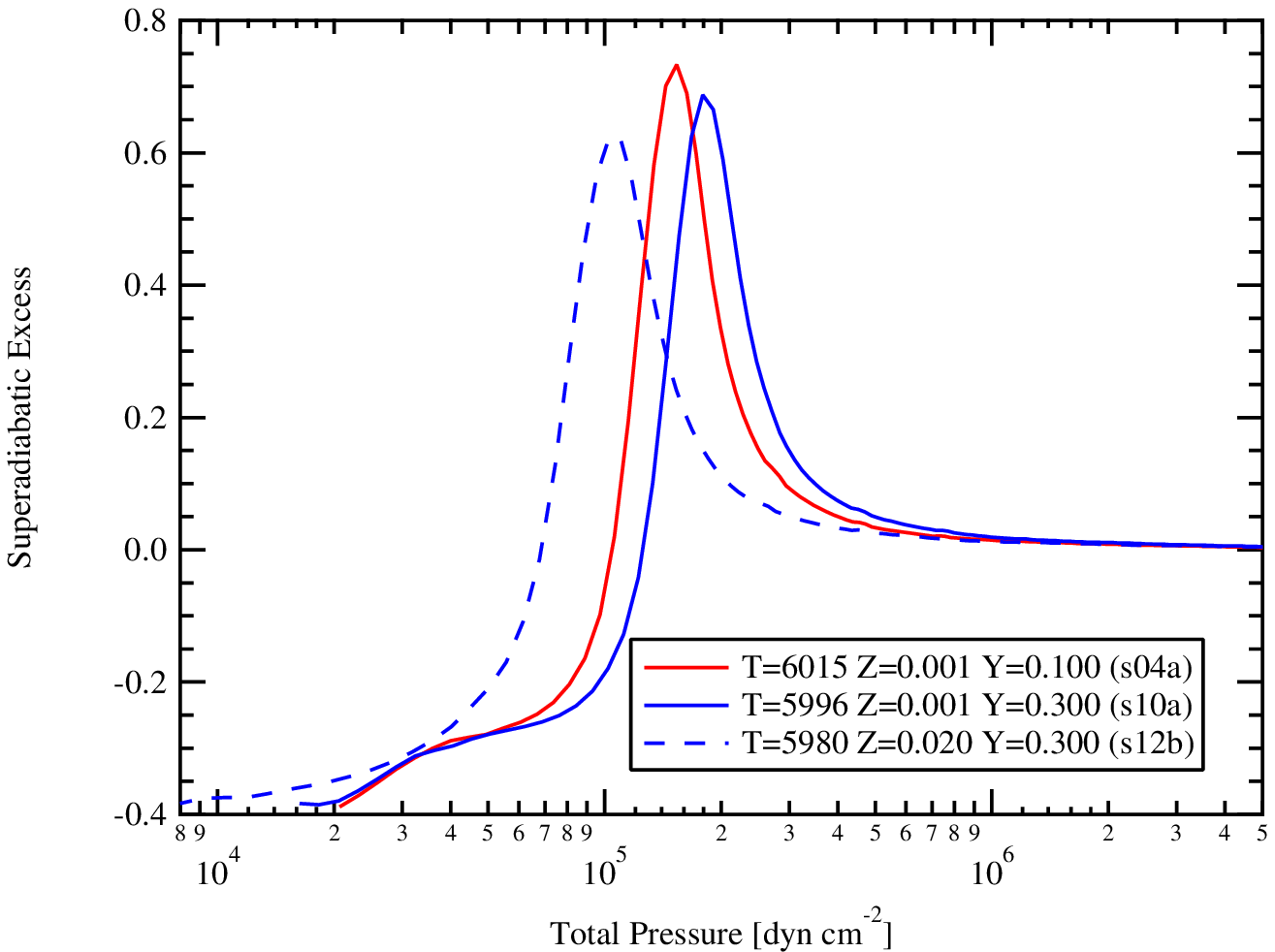}
\epsscale{1.2}
\plotone{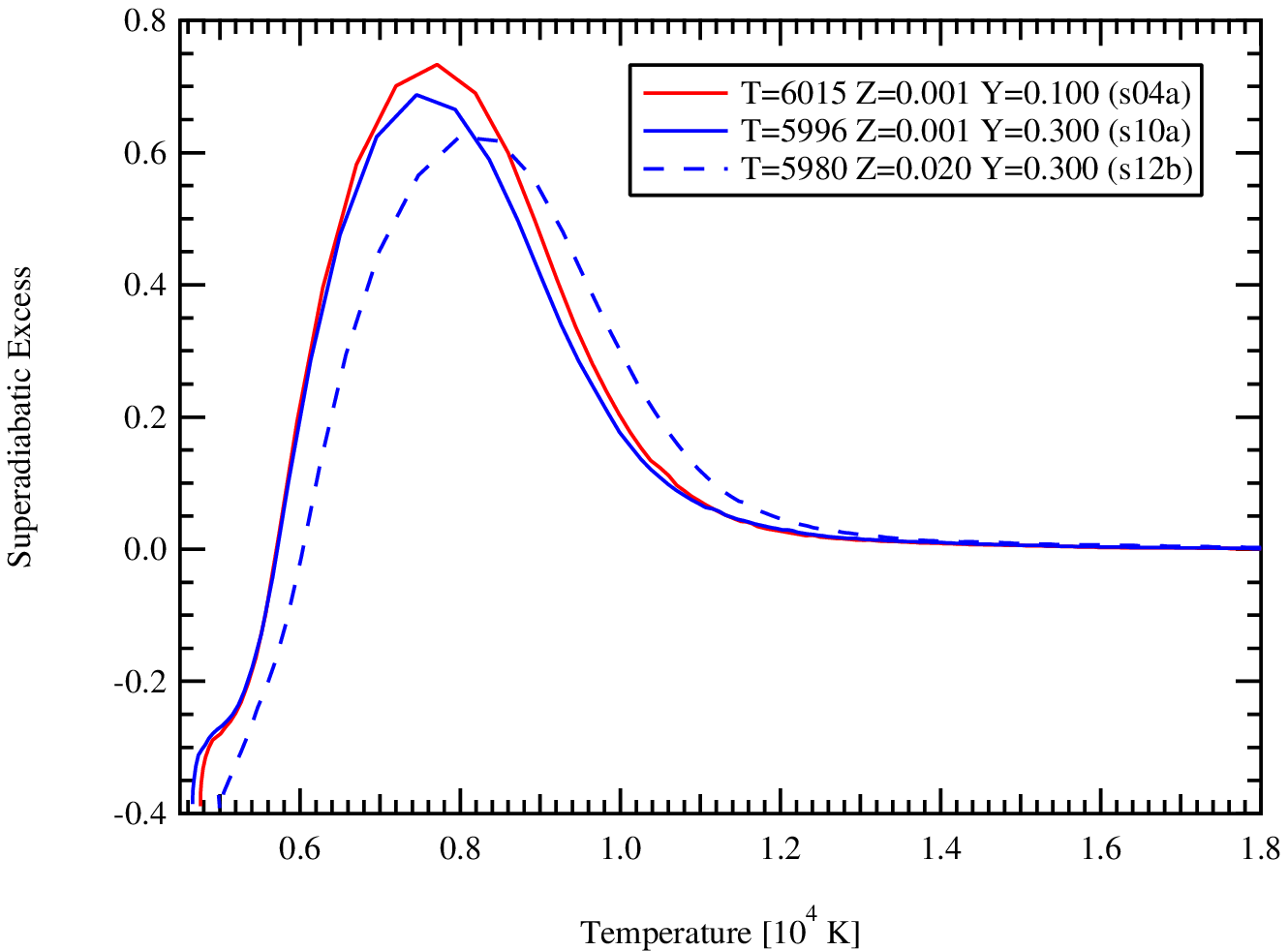}
\end{minipage}
\caption{Superadiabatic excess as a function of total pressure (which includes turbulent pressure) and temperature.  The pressure at the location of maximum superadiabaticity is sensitive to $Y$ and $Z$, while the temperature is mostly sensitive to $Z$.}
\label{fig:sal}
\end{figure}

As a function of temperature, however, the signature of helium abundance differs slightly from that of metallicity.  The lower panel of Fig. \ref{fig:sal} shows that even a large change in $Y$ (increasing $Y$ by a factor of three) does little to change the degree of superadiabaticity, or the temperature at which it is maximal.  An increase of metallicity, on the other hand, pushes the superadiabatic peak to lower pressure, but higher temperature.

The location and extent of the SAL is closely coupled with the hydrogen ionization zone.  {When measured against temperature, the location of the peak of the superadiabatic excess is largely insensitive to helium abundance, which closely mirrors the behavior of the adiabatic gradient.} The location of the hydrogen ionization zone depends on chemical composition, but the lower panel of Fig. \ref{fig:gradad} shows that helium abundance only induces a relatively small adjustment to the temperature of the ionization region.  Metallicity causes a more significant shift in temperature because the electrons with low ionization potential are the source of electron pressure.

The superadiabatic excess is not directly observable, but the differences caused by chemical composition might leave signatures in the optically thin layers.  In particular, differences in the thermal structure of the SAL lead to changes in the convective velocities and turbulent pressure support.  Although convection is not driven in the atmosphere, the velocity field from the convection below is still imprinted on it through convective overshoot.  Convective velocities and overshoot are dealt with in more detail in sections \ref{sec:velocities} and \ref{sec:overshoot}, respectively.

\section{Convection-Zone Dynamics}\label{sec:dynamics}

\subsection{Mean Convective Velocities}\label{sec:velocities}

For a given surface gravity and composition, the strength of the convective velocity field is determined by the energy flux. A star with a higher effective temperature will have faster convective flows in order to carry the increased energy flux which scales as $F \propto \sigma T_{\rm eff}^4$.

Unlike MLT, which represents convection with equal areas of up-flowing and down-flowing material, realistic convection exhibits a strong asymmetry in the velocity field \rev{\citep[e.g.][]{1982ApJ...255..200G, 2009ApJ...697.1032G, 2010ApJ...710.1003G, 2010ApJ...721..670G, 2003MNRAS.340..923R, 1989ApJ...342L..95S, 1998ApJ...499..914S}}.  The buoyant rising fluid is characterized by a larger filling factor through nearly all of the simulation domain \rev{\cite[e.g.][]{2010Ap&SS.328..213T}}.  Upflows maintain nearly constant entropy as they rise almost adiabatically \rev{\citep[e.g.][]{1999A&A...346..111L,1999AAS...194.2104S}} until they reach the SAL near the surface and radiate the energy away.  Near the surface, where radiative losses are strong, the upflows form the characteristic granulation pattern, where they are separated from each other by narrow and fast flowing intergranular lanes that cycle the material back into the convection zone.

Altering the chemical composition of the convection zone will modify the nature of the convective-to-radiative transition and result in a change in convective velocity, the temperature gradient in the atmosphere, and also result in visibly different granulation. \citep[e.g.][]{2013ApJ...767...78T, 2008PhST..133a4004C, 2008MmSAI..79..649C}.  The appearance of low-Z granulation presented in \citet{2013ApJ...767...78T} showed less laminar flow as a result of increased cooling rates from the lower opacity.  Increasing $Y$ has a similar effect on opacity as decreasing $Z$, but the magnitude of the effect is much smaller, so the visual differences in the appearance of the granulation are more subtle.  The consequence of changing $Y$ is more readily apparent in quantitative comparisons.   

\begin{figure*}
\epsscale{1.7}
\begin{minipage}[b]{0.3\linewidth}
\plotone{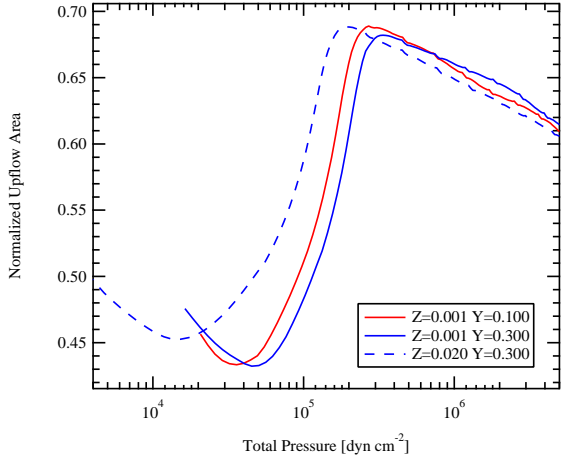}
\end{minipage}
\hspace{3cm}
\begin{minipage}[b]{0.3\linewidth}
\epsscale{1.7}
\plotone{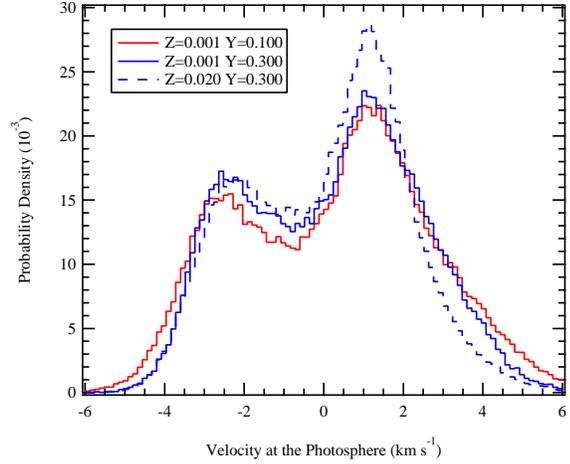}
\end{minipage}
\caption{\textit{Left}: Radial variation of normalized upflow area (filling factor).  The precipitous drop in the upflow area tracks the transition from convective to radiative energy transport. The location of the transition depends on both $Y$ and $Z$. \textit{Right}: The average distribution of photospheric velocities for the same three simulations in the left-hand panel.  Composition appears to affect upflows and downflows differently, with helium and metallicity affecting the strength of the downflows, and the nature of the upflows, respectively.}
\label{fig:areas}
\end{figure*}

The radial variation in normalized upflow area as well as the photospheric velocity distributions in Fig. \ref{fig:areas} show an adjustment in response to changes in $Y$ and $Z$.  The upflowing gas rises nearly isentropically \rev{\citep[e.g.][]{1999A&A...346..111L,1999AAS...194.2104S}} through most of the convective envelope, until it approaches the surface and can begin to radiate.  As energy is radiated away, the gas cools and is then cycled back into the convective envelope.  The relative surface area of the hot, rising granules is sensitive to the radiative cooling rate near the surface \rev{\citep[e.g.][]{2013ApJ...767...78T, 2007A&A...469..687C, 1999A&A...346L..17A}}.  A high cooling rate (low opacity) permits the granules to begin radiating energy sooner, so the drop in upflow area occurs deeper in the star, which is shown in the left panel of Fig. \ref{fig:areas}.  The right hand panel presents the velocity distribution from the photosphere of the same three simulations.  

The bimodal nature from the distinct upflows and downflows is prominantly displayed in the velocity distribution.  \rev{The histogram is qualitatively similar to those of \citet{1998ApJ...499..914S},  \citet{2000SoPh..192...91S}, and \citet{2000A&A...359..669A}, but the three profiles in Fig. \ref{fig:areas} show the effect of composition}.  The smaller peak at negative velocity corresponds to the downflows, which are faster than the upflows but occupy less area.  The larger peak at positive velocity corresponds to the rising granules.  

\rev{The nature of the downflows does not appear to be strongly affected by metallicity, particularly in the high velocity wing of the probability distribution.  Downflows are the consequence of mass conservation, and exist so that the gas can be cycled back into the convective envelope.  Radiative cooling rates from opacity are not as important in the downflows, so changes to $Z$ introduce a small effect.  The downflow component of the velocity histogram, however, does appear to be sensitive to helium abundance.  Increasing $Y$ substantially increases the mean molecular weight (Fig. \ref{fig:mmwt}), particularly in the atmosphere.  Since the mean molecular weight in the atmosphere is not significantly altered by increasing $Z$, this suggests that the nature of the intergranular lanes depends to some degree on the helium abundance. }

Metallicity influences the upflows more than the downflows.  The second peak in the distribution is considerably larger in the high-$Z$ simulation, which means that the granules are occupying a greater surface area.  This is consistent with the larger and more laminar visual appearance of the high-$Z$ granules in \citet{2013ApJ...767...78T}.  These differences in the velocities are areas could have an effect of the shape of spectral lines, which we examine in more detail in Section \ref{sec:spectral}.

\begin{figure}
\epsscale{1.2}
\begin{minipage}[b]{1.0\linewidth}
\plotone{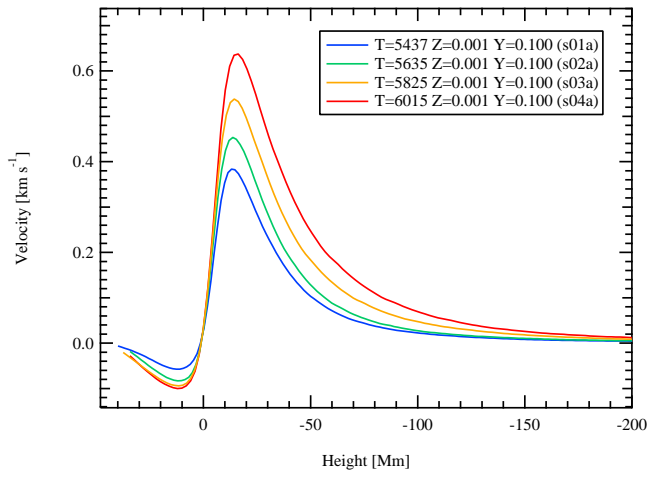}
\epsscale{1.2}
\plotone{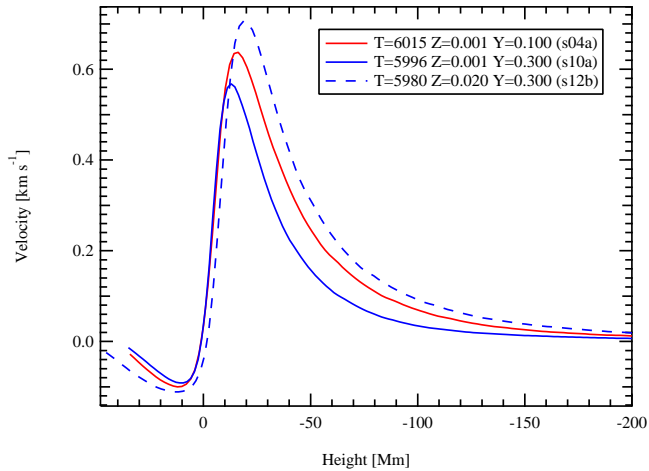}
\end{minipage}
\caption{Mean convective velocity as a function of height.  The zero point of the height axis is set to the photosphere.  Convective velocities are clearly sensitive to the energy flux (top), with higher effective temperature requiring more vigorous convection to transport the energy.  Composition (bottom) is an important factor as well, with changes to $Z$ resulting in a larger shift than changes to $Y$.}
\label{fig:velocity}
\end{figure}

If upflows are defined as having a positive velocity, the large filling factor results in a positive spatial average through most of the domain.  \rev{This is a general characteristic of sub-photospheric stellar convection, and the role of the filling factor and its implications for modeling convection has been discussed in the literature \citep[e.g.][]{1998ApJ...493..834C, 1999ApJ...526L..45K, 1999ASPC..173..157K}}. The situation reverses above the photosphere where convection is no longer driven and overshoot occurs \rev{\citep[e.g.][]{1990A&A...228..155N, 2005MNRAS.362.1031R}}, and the mean velocity is negative.  Fig. \ref{fig:velocity} presents the spatially averaged vertical velocity as a function of height.  The zero point of the height axis has been set to the photosphere, where $\langle T \rangle=T_{\rm eff}$.

The velocity reaches a maximum very near the peak of the superadiabatic excess, and drops sharply approaching the photosphere from below.  For a given chemical composition and surface gravity, the maximum mean velocity scales with energy flux, \rev{which is consistent with the simulation grids of  \citet{2013ApJ...769...18T} and \citet{2013arXiv1302.2621M}}.  The upper panel of Figure \ref{fig:velocity} compares the velocity profiles from a set of four simulations with the same composition.  The three simulations in the lower panel, however, have the same energy flux, so differences in the velocity field are caused by changes to the helium or metal element abundances.

Simulations with higher effective temperatures require more vigorous convection to carry the energy flux.  For the simulation set presented in the top panel of Figure \ref{fig:velocity}, increasing the effective temperature by 10\% results in mean convective velocities that are 60\% larger.  The convective velocities are shown to increase with opacity, which is larger with increasing metallicity and decreasing helium.  The lower panel of Figure \ref{fig:velocity} shows a 25\% increase in mean convective velocity from increasing $Z$ from $0.001$ to $0.020$.  \rev{As reported by \citet{2013ApJ...767...78T}, the} larger velocities are a direct result of the decreased density in the lower-opacity simulations, wherein the less dense gas must mix more vigorously in order to maintain the energy flux.  The temperature at the height of the maximum velocity shows a similar sensitivity to composition as that of the maximum superadiabaticity (Fig \ref{fig:sal}).  The temperature at this location increases with metallicity, but is largely insensitive to helium abundance. 

\begin{figure}[h]
\epsscale{1.20}
\plotone{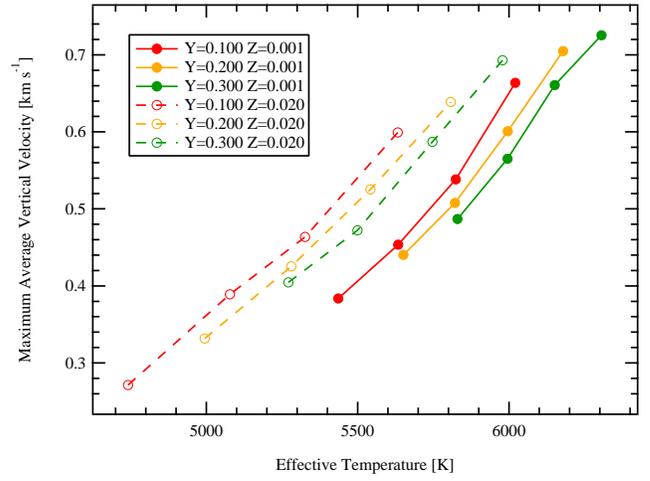}
\caption{Convective strength as measured by the average maximum vertical velocity as a function of effective temperature.  Each data point is extracted from a simulation, and simulations with a particular composition are connected with lines.  Metallicity seperates the simulations into two distinct groups, while helium abundance introduces a smaller offset.}
\label{fig:velocity_trends}
\end{figure}

We present the range of maximum vertical velocity over the grid of simulations Fig. \ref{fig:velocity_trends}.  Each point corresponds to the peak of the velocity profiles from the corresponding simulation (the maximum value from velocity profiles similar to those in Fig. \ref{fig:velocity}).  Data points from simulations that share a chemical composition are connected with lines.  There is a strong visible trend with energy flux, with the higher {\teff} simulations exhibiting larger convective velocities.  Chemical composition introduces an offset in the $\langle w \rangle$-$\langle T_{\rm eff} \rangle$ trends.  Changing metallicity, which has a strong  effect on opacity, induces a larger shift than the helium abundance.  The magnitude of the effect of helium on the vertical velocity appears to be generally insensitive to metallicity.  The range in maximum vertical velocity as a function of helium abundance at a particular {\teff} is essentially the same in the high ($Z=0.020$) and low ($Z=0.001$) metallicity simulations. 

\subsection{Turbulent Convective Velocities}\label{sec:turbulent}

One of the advantages that 3D simulations have over stellar models is the inclusion of turbulent pressure.  It arises automatically and self-consistently in 3D simulations from the convective gas dynamics, and is undoubtedly an important factor in  understanding the structure near stellar surfaces.  

We extract turbulent pressure from the density, $\rho$, and RMS vertical velocity, $w_{\rm rms}$.  Defined as
\begin{equation}\label{eqn:pturb}
P_{\rm turb} = \rho w_{\rm rms}^2 ,
\end{equation}

the turbulent pressure is most important where the superadiabatic excess is large.  Gas pressure dominates below the SAL, and convective velocities are weak above.  Through the SAL, turbulent pressure contributes significantly to hydrostatic equilibrium, reaching values in excess of 15\% of the gas pressure \rev{\citep{1999A&A...351..689R, 2009MmSAI..80..701K, 2012A&A...539A.121B}}.  

\rev{Recent systematic studies of near-surface stellar convection show that turbulent pressure varies as a function of effective temperature and surface gravity \citep{2013ApJ...769...18T} as well as metallicity \citep{2013ApJ...767...78T, 2013arXiv1302.2621M}.  In this section we examine the variation of turbulent pressure with composition, including the effect of changing metallicity and helium abundance.  The variation in maximum turbulent pressure is consistent with previously published work of \citet{2013ApJ...767...78T}, and qualitatively similar to the values reported by \citet{2013ApJ...769...18T} and \citet{2013arXiv1302.2621M}.}

\begin{figure}[h]
\epsscale{1.20}
\plotone{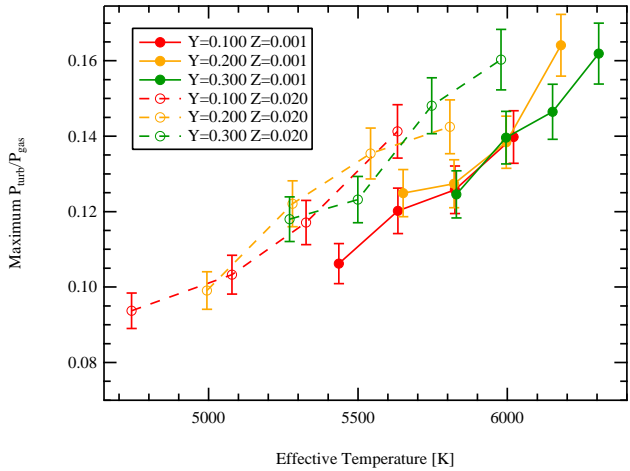}
\caption{Similar to Fig. \ref{fig:velocity_trends} but showing maximum turbulent pressure relative to gas pressure.  Turbulent pressure increases with both {\teff} and $Z$, but shows little remarkable sensitivity to helium.}
\label{fig:pturb_trends}
\end{figure}

The fraction of turbulent pressure support depends on composition, as demonstrated in Fig. \ref{fig:pturb_trends}, which presents the maximum turbulent pressure (relative to gas pressure) for all the simulations in the grid.   Similar to the trend in velocity, the largest effect is from varying the energy flux, and the contribution from turbulent pressure is quite sensitive to metallicity as well.  Quite surprisingly, there is little remarkable change induced by changes to helium abundance.  Our simulations show sensitivity to helium, but it is much smaller than the effect of metallicitiy.

\subsection{Overshoot}\label{sec:overshoot}

Chemical composition strongly influences the convective dynamics and thermal structure across the SAL, where the convection is most turbulent.  This region is below the photosphere and not directly observable, but momentum carries the convective dynamics above the convective envelope and into the convectively stable radiative atmosphere.  This so-called convective overshoot may well bear the signature of chemical composition.  Indeed, using 3D simulations in conjunction with spectral synthesis has proven to be a useful method for accounting for macro-turbulent broadening and produces realistic spectral line profiles \citep[e.g.][]{2000A&A...359..729A}. 

We estimate the strength, or length scale, of convective overshoot by measuring correlations between thermodynamic and convective properties. When convection is coherent, fluctuations in temperature and velocity will be positively correlated.  This is simply because hot regions rise (with positive velocity) and cool regions sink (with negative velocity).  

The correlation function between two turbulent quantities is defined as:
\begin{equation}
\label{eqn:corr}
C[q_1,q_2] = \frac{\langle{q_1 q_2}\rangle - \langle{q_1}\rangle \langle{q_2}\rangle}{q_{\rm 1,rms}q_{\rm 2,rms}} .
\end{equation}

where the RMS of a quantity is calculated using:
\begin{equation}
\label{eqn:rms}
q_{\rm rms} = \sqrt{\langle{q^2}\rangle - \langle{q}\rangle^2} .
\end{equation}

The correlation function will be near unity where convection is efficient, and will fall in regions near and above the photosphere.  This provides a way to define the edge of the overshoot region; namely, where the correlation function reaches zero.  Additionally, the limit of driven convection can be defined as where superadiabatic excess goes to zero.  Because of convective overshoot, these two boundaries will not be at the same location (the overshoot boundary will be above the driven convection boundary) in the SAL, and the overshoot length is defined as the difference between them.

\begin{figure*}[h]
\epsscale{1.0}
\plotone{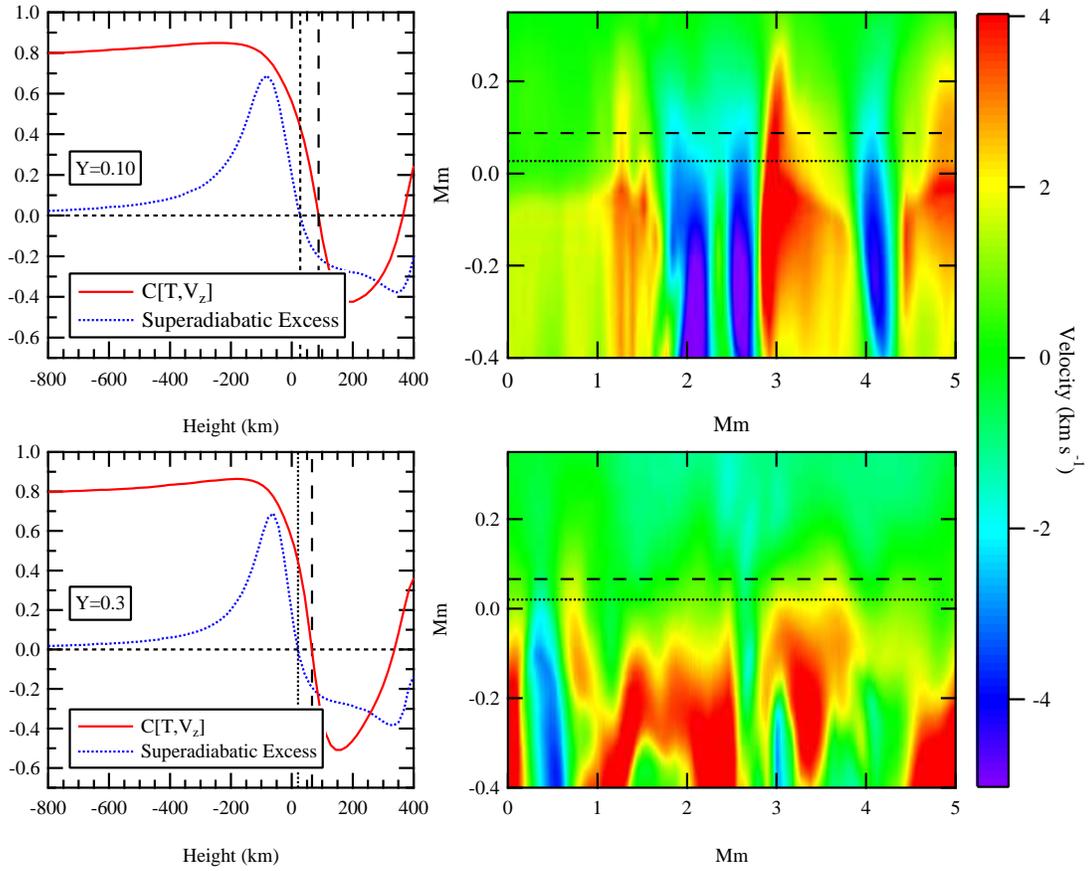}
\caption{Overshoot length for two simulations with varied helium abundance.  Overshoot length is as measured by the seperation between where the superadiabaticity and the correlation function of turbulent quantities go to zero.  The effect of helium abundance on overshoot is markedly smaller than that of metallicity, as reported by \citet{2013ApJ...767...78T}.}
\label{fig:overshoot_image}
\end{figure*}

\begin{figure}[h]
\epsscale{1.20}
\plotone{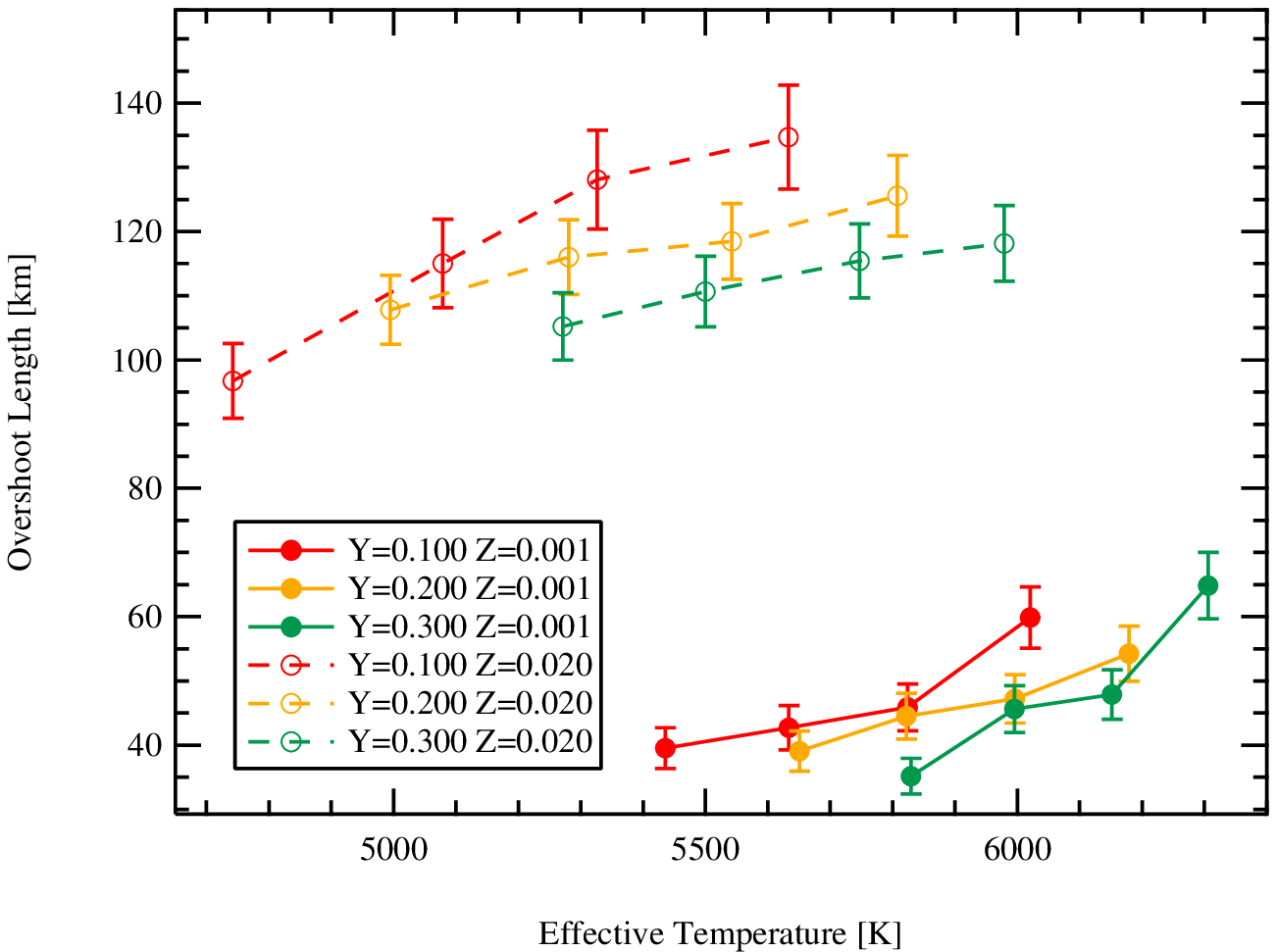}
\caption{Similar to Fig. \ref{fig:velocity_trends} but showing overshoot length.  The extent of the overshoot follows the trends in convective velocitiy, inreasing with larger {\teff} and $Z$.  The effect of $Y$ is smaller, but significant.}
\label{fig:overshoot_trends}
\end{figure}

An example of the effect of helium abundance on overshoot is provided in Fig. \ref{fig:overshoot_image}.  Panels on the left side of the figure show the horizontally and temporally averaged radial profiles of superadiabaticity and the turbulent correlation functions that define the overshoot length.  The right-hand panels show vertical slices of the velocity field from a simulation snapshot.  The velocities in the low-$Y$ simulation are larger, which is consistent with Figs. \ref{fig:velocity} and \ref{fig:velocity_trends}, and the distance (extending higher into the atmosphere) over which the velocites remain coherent is visbily larger, resulting in a larger space between height where $\nabla-\nabla_{\rm ad}=0$ and $C[T,V_z]=0$.  The dashed lines on the left are also displayed on the vertical slice for reference.  The difference in overshoot is smaller than what was observed in the metallicty comparison of \citet{2013ApJ...767...78T}, but it is nonetheless apparent.  We present the overshoot length for the entire grid of simulations in Figure \ref{fig:overshoot_trends} in a similar manner to Figs. \ref{fig:velocity_trends} and \ref{fig:pturb_trends}.  The amount of overshoot into the atmosphere is correlated with the mean convective velocities.  The upflows in simulations with larger velocities (hotter {\teff}, larger $Z$, or smaller $Y$) have more momentum, and so they remain as coherent structures over a greater distance.

\section{Spectral Line Shapes}\label{sec:spectral}

We have shown in Section \ref{sec:dynamics} that the mean velocity, filling factor, photospheric velocity distribution, and overshoot length scale are sensitive to helium and metallicity.  The composition-induced differences in convective dynamics suggests that Doppler broadening of spectral lines may bear the signature of helium abundance.  Forming over a range of depths in the atmosphere, the turbulent broadening of spectral lines is complex, and depends on the radial variation of velocity, as well as thermal structure and filling factor from flow asymmetry.  The large, hot upflows will contribute the most to spectral features, and differences in the velocity field between the photosphere and higher atmospheric layers will change the way that line wings and cores are affected.

In this section we describe a simple ‘toy’ model for spectral line formation, and use it to estimate the kind of effect that helium abundance might introduce. We emphasize that this is a simple model, and this analysis should be followed up by more sophisticated 3D spectral synthesis \citep[e.g.,][]{2008psa..conf..133L, 2013arXiv1303.2016P}.

\subsection{A Simple Model}

Our spectral model is derived from the weak-line treatment described in \citet{1989isa2.book.....B}, and inspired by the modeling technique of \citet{1990A&A...228..218D}.  

We model a spectral line as the cumulative contribution from lines originating from different columns that span the horizontal extent of the simulation domain.  This is qualitatively similar to the model of Dravins, except that instead of a four-component approach, we have one component for each column. We resample each simulation snapshot to 64x64 columns, which is sufficient to capture the effect of the bimodal velocity field. The line profile from each component is Doppler shifted according to the simulated velocity field.  The model is applicable only to the stellar disk center, and does not include effects such as limb-darkening.

The contribution to the spectral line from a particular column is represented by a Voigt profile. The function is characterized by a damping parameter ($a$) and a frequency offset ($u$), which is defined as:
\begin{equation}
u = \frac{\nu-\nu_0}{\Delta \nu_D} ,
\end{equation}
where $\Delta \nu_D$ is the Doppler width:
\begin{equation}
\Delta \nu_D = \frac{v_0}{c}\sqrt{\frac{2kT}{m}}.
\end{equation}

The intrinsic Doppler width of the Voight profile depends on the atomic mass of the element, which we set to $56$ to correspond to iron. The damping parameter is set to $0.001$ which is consistent with the `weak' lines in \citet{1990A&A...228..218D}.

The Voigt profile is Doppler shifted as a function of wavelength according to the vertical velocity from the column in the simulation, at the formation depth of the line.  Line wings are formed closer to the photosphere where the velocities tend to be higher, and line cores are formed higher in the atmosphere where the velocity is typically smaller.   

The strength of the line is computed using the description provided by \cite{1989isa2.book.....B}, which is valid for unsaturated lines.  The depth of the line relative to the continuum is characterized by a parameter $R$, which is related to the equivalent width.  \rev{Larger values of $R$ create stronger lines with cores that form higher in the atmosphere. We keep relatively small values of R (which correspond to weak spectral lines) so that we can apply the weak-line treatment of  \citet{1989isa2.book.....B} and minimize potential effects from the simulation boundary.}

The line depth is: 
\begin{equation} \label{eqn:r}
R = \frac{F_C - F_\lambda}{F_C} = \frac{B_\lambda (\tau_c=2/3) - B_\lambda (\tau_\lambda=2/3)}{B_\lambda (\tau_c=2/3)} ,
\end{equation}
where $F_C$ and $F_\lambda$ are the continuum and lines fluxes, respectively, which under local thermodynamic equilibrium are given by the Planck function.  \rev{The wavelength dependence of the Doppler shift is introduced indirectly through the temperature gradient in the simulation.  For a particular column in the simulation, we determine the depth of the line formation by solving Equation (9) for $B_\lambda(\tau_\lambda=2/3)$ with a given value of $R$ and $B_\lambda(\tau_c=2/3)$.  The Planck function, which depends only on temperature for a given wavelength, is then used to determine the temperature of the layer at which the line is formed, which in turn gives us the depth of formation.  The value of R varies as a function of wavelength from the line wings to the line core, which translates to different formation depths.  The Doppler shift is applied using the velocity field at the depth determined by Equation (9)}.

The contribution from bright (blueshifted) upflowing granules is significantly larger than that from a cool (redshifted) downflowing intergranular lanes.  The combination of the large filling factor for upflows and temperature contrast between upflows and downflows means that they are both brighter and larger.  The upflow contributes much more to the final spectrum, but the downflow has a larger Doppler shift relative to line center, and in the opposite direction.  Furthermore, the lower continuous opacity in the downflows causes the line of sight to penetrate deeper in the atmosphere where the velocities are stronger.  The final line profile is a sum of profiles over the horizontal surface of the simulation, weighted by the continuum flux.  We use several simulation snapshots over several hours of simulated time to produce the final time-averaged profile.

\subsection{Helium Abundance and Spectral Line Shapes}
\label{sec:speclines}

Insofar as they are affected by the photospheric velocity field from convection, the basic properties of spectral line shapes are defined by two principal factors.  These are the asymmetry of the flow (with upflows forming the granule surfaces and having a large filling factor near the photosphere), and the radial velocity profile.  

The vertical velocity profile from a particular column from a simulation snapshot determines the Doppler shift at a particular depth. The line wings are formed close to the photosphere, and so they bear a strong imprint of the photospheric granulation.  Line cores form higher in the atmosphere where the velocities are smaller.  The magnitude of the Doppler shift applied the core relative to the wings depends on the velocity gradient, and the strength of the line.  Cores of stronger lines form higher in the atmosphere than those of weaker lines, so the Doppler shift is smaller. 

To rough approximation, the absolute shift in line cores will be determined by the upflows (see, for example, \citet{graybook} which describes a simple two-stream model for visualizing granulation).  Even accounting only for the radial velocity profile sets the stage for explaining the absolute Doppler shift of lines, or third signature of convection \citep{2009ApJ...697.1032G}.  Asymmetry in the flow, however, also contributes to the shape of the line.  Lines formed in the cool, fast intergranular lanes will be strongly redshifted, but lower intensity relative to the upflows.  

\begin{figure*}[h]
\epsscale{1.00}
\plotone{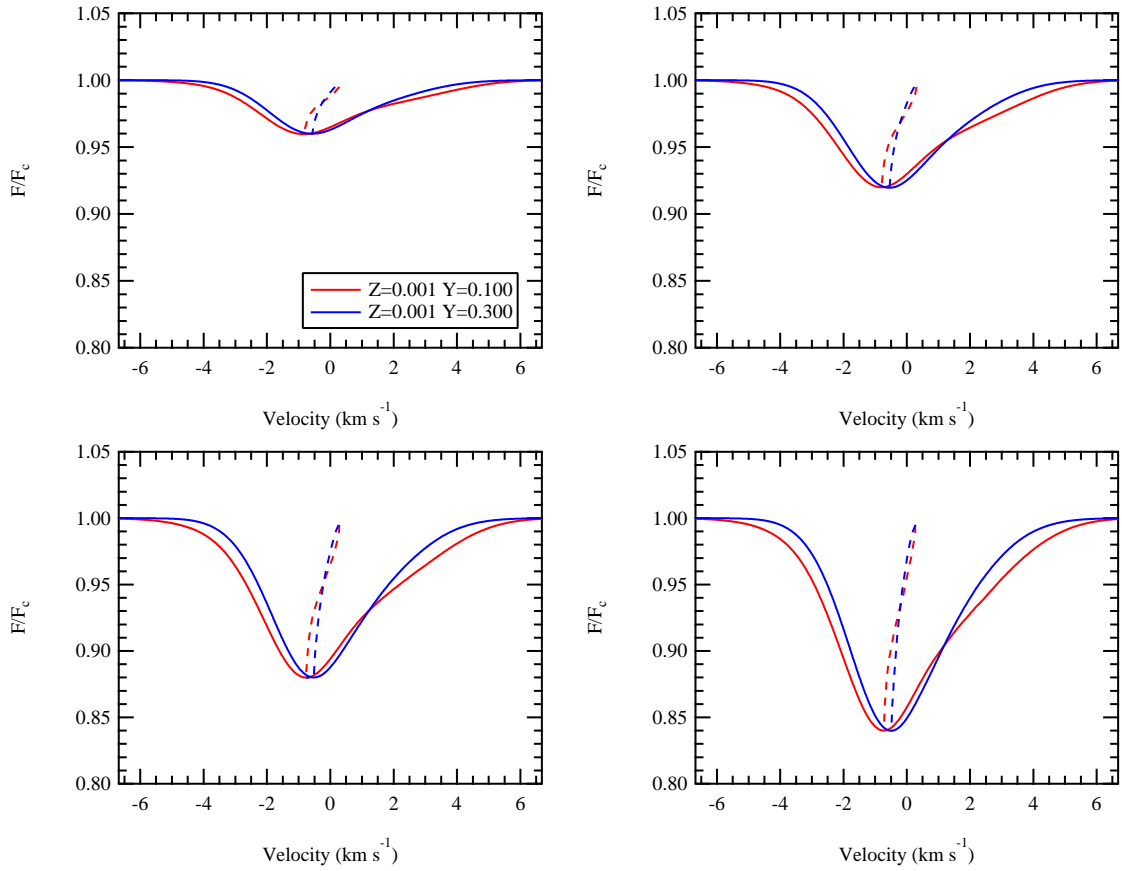}
\caption{Comparing the effect of helium abundance on weak spectral line profiles for four line strengths.  In each panel, high and low helium abundance are identified with blue and red lines, respectively. Helium abundance clearly affects the line profile.  In particular, the higher velocities in the low-$Y$ simulation results in increased Doppler shifts and greater curvature to the line bisector.}
\label{fig:lines}
\end{figure*}

The combined effect of the velocity profile and the flow asymmetry explains the typical `C' shaped line bisector \citep[e.g.][]{graybook, 2010ApJ...710.1003G}.  We show the line shapes as a function of line strength in the upper panels of Fig. \ref{fig:lines}.  The location of the line cores is most strongly shifted in the weak lines, where the formation depth is closest to the photosphere and the velocities are high.  The line core becomes progressively less blueshifted as the depth of the line grows.  Weak lines also tend to show the upper half of the `C', as their formation depths do not extend very high into the overshoot layer.  Stronger lines with deeper cores begin to complete the `C' shape.

Fig. \ref{fig:lines} compares two line bisectors computed from simulations with different helium abundances.  Each panel in the figure compares the same two helium abundances, but at different line strengths.  The change in velocity field as a result of the different helium abundances induces a corresponding adjustment to the shape of the bisector, with the larger velocities (and steeper gradient) in the low-$Y$ simulation (Fig. \ref{fig:velocity}) causing the larger shift in the line.

\begin{figure}[h]
\epsscale{1.20}
\plotone{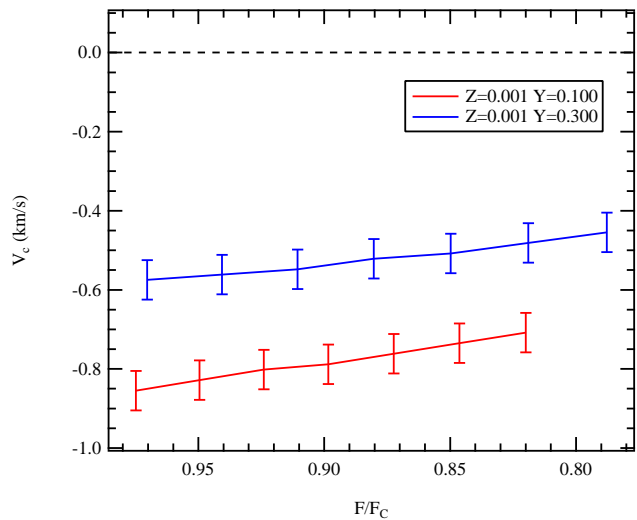}
\caption{Absolute Doppler shifts (of line cores) as a function of line strength.  The line shift decreases with line strength because the cores of deeper lines are formed higher in the atmosphere where the velocities are weaker.}
\label{fig:lineshift}
\end{figure}

After extracting the Doppler shift of the line cores, we present it in Fig. \ref{fig:lineshift} as a function of line strength, which is essentially a third signature plot.  Each curve traces the velocity of the line core as a function of line strength from a set of spectral lines similar to those presented in Fig. \ref{fig:lines}.  The systematically larger shifts in the low-$Y$ simulation are a consequence of the larger velocities and overshoot length discussed in Sections \ref{sec:velocities} and \ref{sec:overshoot}.  

The line profiles computed in our simple model suggest that spectral line shapes might bear the signature of helium abundance.  We stress that this result is from a simple model, and should be followed up with spectral synthesis.  Furthermore, extracting information about helium composition in this manner is only possible if the metallicity is known.

\section{Alpha-Enhancement in Globular Cluster Environments}\label{sec:alpha}

Our grid of simulations is divided into those with solar-like metallicity of $Z=0.020$ and those that are metal-poor with $Z=0.001$. Old metal-poor stellar populations, such as those found in globular clusters, are often enriched in $\alpha$-elements.  It is expected that the details of the mixture of metals will have a considerably smaller effect on convection than the metal mass fraction, but the effect may be detectable in the simulations.  Given that properties of convection are considerably more affected by changes to $Z$ than $Y$, we are interested in measuring the magnitude of the effect on convection that might be a result of the details of the heavy element mixture.  Since the primary source of opacity is from the H$^-$ ion, changing the number of ionized electrons by altering the heavy element mixture could have a measurable effect on the structure and dynamics of the SAL.  As described in section \ref{sec:microphysics}, the ionized electrons from the metals contribute to the electron pressure, which alters the hydrogen ionization zone, and in turn affects the SAL structure. 

Examples of particular interest include the stellar populations of $\omega$ Centauri and NGC 2808, both of which show signs of enriched helium abundance \citep{2005ApJ...621L..57L}.  We use 3D simulations to determine to what degree $\alpha$-enhancement affects near-surface convection.  In this section we summarize results from a tests using a grid of simulations that is separate from the grid described in previous sections.  These simulations also have a fixed surface gravity, and span a range of effective temperature near the main sequence turnoff of $\omega$ Centauri.  Properties of this grid are presented in Table \ref{tab:alpha}. In a manner similar to the grid defined in Section \ref{sec:grid}, the simulations are broken into sets corresponding to different compositions.  Instead of changing the helium mass fraction, the alpha enhancement is altered by using the enhanced tables of \citet{2005ApJ...623..585F} for the mixture of \citet{1998SSRv...85..161G}.   The ratio of hydrogen-to-metal mass fraction is kept fixed.

\begin{table*}[t]
  \centering
  \caption{Properties of the simulations with varied $\alpha$-element enhancement.  All simulations have the same metallicity, hydrogen mass fraction, and surface gravity ($\log g = 4.30$).}
  \label{tab:alpha}
  \begin{center}
    \leavevmode
    \begin{tabular}{cccccccc} \hline \hline              
[$\alpha$/Fe] &  $\log(T_{\rm eff})$     & Z   & X & $\Delta x \times \Delta z$ (km) & $N_x \times N_z$ & $\Delta X$ (Mm) & $\Delta Y$ (Mm) \\ \hline 
 0.00 & 3.791   & 0.00057 & 0.870 & $84.94 \times 25.80$ & $95 \times 225$ & 8.07 & 5.80 \\
 0.00 & 3.794   & 0.00057 & 0.870 & $84.94 \times 25.80$ & $95 \times 225$ & 8.07 & 5.80  \\
 0.00 & 3.796   & 0.00057 & 0.870 & $84.94 \times 25.80$ & $95 \times 225$ & 8.07 & 5.80  \\
 0.00 & 3.799   & 0.00057 & 0.870 & $84.94 \times 25.80$ & $95 \times 225$ & 8.07 & 5.80  \\
 0.00 & 3.802   & 0.00057 & 0.870 & $84.94 \times 25.80$ & $95 \times 225$ & 8.07 & 5.80  \\
 0.80 & 3.795   & 0.00057 & 0.870 & $84.94 \times 25.80$ & $95 \times 225$ & 8.07 & 5.80  \\
 0.80 & 3.797   & 0.00057 & 0.870 & $84.94 \times 25.80$ & $95 \times 225$ & 8.07 & 5.80  \\
 0.80 & 3.800   & 0.00057 & 0.870 & $84.94 \times 25.80$ & $95 \times 225$ & 8.07 & 5.80  \\
 0.80 & 3.802   & 0.00057 & 0.870 & $84.94 \times 25.80$ & $95 \times 225$ & 8.07 & 5.80  \\
 0.80 & 3.805   & 0.00057 & 0.870 & $84.94 \times 25.80$ & $95 \times 225$ & 8.07 & 5.80  \\ \hline
    \end{tabular}
  \end{center}
\end{table*}

\begin{figure}[h]
\epsscale{1.20}
\plotone{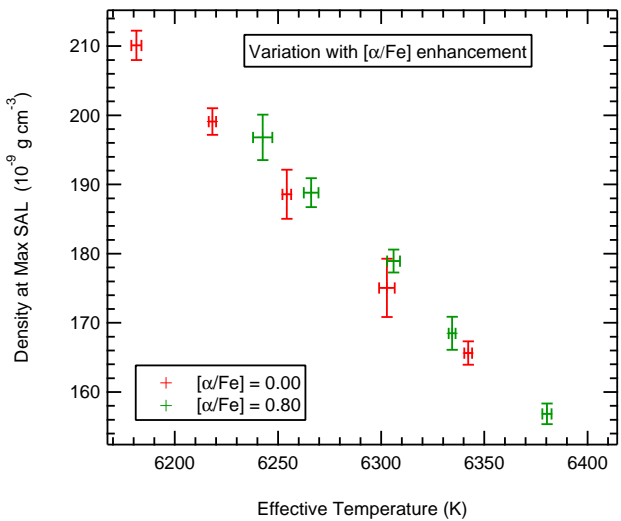}
\caption{Density at the location of maximum superadiabaticity in simulations with normal and $\alpha$-enhanced GS98 abundances.  The effect of changing the heavy element mixture is considerably smaller than changing $Z$ or $Y$, as seen in Fig. \ref{fig:rhovsp}.}
\label{fig:alpha}
\end{figure}

Relative to changes in the $Z$ and $Y$ mass fractions, the mixture of heavy metal elements introduces a rather minor effect on the convective dynamics.  Although the effect is quite small, changing the $\alpha$-element abundance has some detectable consequences in the mean quantities, which have smaller statistical errors than quantities computed from turbulent fluctuations.  At these low metallicities, enhancing the $\alpha$-elements adjusts the opacity slightly, resulting in a small alteration in density through the SAL.  Fig. \ref{fig:alpha} compares the density at the location of the peak superadiabaticity for the normal and $\alpha$-enchanced GS98 tables.  

\rev{Although the changes induced by $\alpha$-element enhancement to the density through the SAL is detectable, it is considerably smaller than the effect of changing metallicity or helium abundance.   For example, the increase in density at a particular effective temperature is approximately $25$ and $40$ times smaller than the range in SAL density as a function of helium and metallicity in Fig. \ref{fig:rhovsp}, respectively.}  As a consequence of changing the density stratification by enhancing the $\alpha$-element abundance, the convective velocities would need to change to maintain a fixed energy flux.  The effect on the convective velocities is small enough that we cannot measure it to a level of good statistical significance.

Higher-order quantities such as the turbulent pressure and adiabatic gradient have larger statistical error bars, and variations between simulations with different alpha enhancements are within the statistical uncertainty of the measurement.  This is not a surprise considering that changing the metal element mass fraction itself introduced no remarkable change to the superadiabatic excess.  While the results from \citet{2013ApJ...767...78T} demonstated that turbulent pressure is sensitive to changes in metallicity, our current set of alpha-enchanced simulations suggest that any adjustment caused by enhancing the alpha elements appears to be below the precision of the simulations.  The precision of the quantities measured from simulations can be improved by gathering statistics over a much longer baseline, but we do not pursue this further since the effect is expected to be small.

\section{Helium Abundance and Surface Gravity}\label{sec:gravity}

\rev{In classical stellar atmospheres, the effect of helium can alter the spectroscopically deduced surface gravity.  As pointed out by, e.g. \citet{graybook}, helium and surface gravity behave in a similar way.  Because the effect of helium and gravity are similar, at least on the mean stratification away from the SAL, it is worth examining to what extent {\logg} and $Y$ have a similar effect on 3D RHD simulations.  In sections \ref{sec:microphysics} through \ref{sec:speclines} we presented various effects of changing the helium abundance on the structure of the SAL, by comparing simulations with different compositions at fixed {\logg} and over a range of {\teff}.   Other published work \citep[e.g.][]{2013ApJ...769...18T, 2013arXiv1302.2621M} has dealt with the variation of convective properties across much of the HR diagram, but here we examine the effect of surface gravity in the context of helium abundance.  We compare the behavior of {\logg} and $Y$ in simulations to determine if they behave similarly even in a 3D turbulent atmosphere.}

\rev{Table 3 summarizes the set of simulations used for this comparison.  The simulations are divided into two groups that overlap in their respective {\teff} ranges.  The two sets have {\logg}$=4.30$ and {\logg}$=4.50$, respectively.  All of the simulations have exactly the same microphysics.  We can directly compare the properties of simulations s02c and s03d, which have identical compositions and essentially the same {\teff}.    These two simulations differ in composition to those presented in previous  sections, so direct comparison is not possible.  We can, however, compare the response of {\logg} to that of $Y$ to see if the effect is in the same direction.  }

\begin{table*}[t]
  \centering
  \caption{Properties of the simulations with varied surface gravity.  All simulations have the same metallicity and  hydrogen mass fraction.}
  \label{tab:logg}
  \begin{center}
    \leavevmode
    \begin{tabular}{ccccccccc} \hline \hline              
  ID & {\logg} & $\log (T_{\rm eff})$ & Z & X & $\delta x \times \delta z$ (km) & $N_x \times N_z$ & $\Delta x$ (Mm) & $\Delta z$ (Mm)  \\ \hline 
 s01c & 4.30 & 3.710 & 0.020 & 0.735 & $50.23 \times 16.06$ & $95 \times 215$ & 4.77 & 3.46 \\
 s02c & 4.30 & 3.729 & 0.020 & 0.735 & $53.56 \times 17.13$ & $95 \times 215$ & 5.09 & 3.68 \\
 s03c & 4.30 & 3.752 & 0.020 & 0.735 & $57.88 \times 18.52$ & $95 \times 215$ & 5.50 & 3.98 \\
 s04c & 4.30 & 3.770 & 0.020 & 0.735 & $63.33 \times 20.26$ & $95 \times 215$ & 6.02 & 4.36 \\
 s01d & 4.50 & 3.687 & 0.020 & 0.735 & $29.96 \times 9.58$ & $95 \times 215$ & 2.85 & 2.06 \\
 s02d & 4.50 & 3.707 & 0.020 & 0.735 & $31.35 \times 10.02$ & $95 \times 220$ & 2.98 & 2.21 \\
 s03d & 4.50 & 3.727 & 0.020 & 0.735 & $33.20 \times 10.62$ & $95 \times 220$ & 3.16 & 2.34 \\
 s04d & 4.50 & 3.748 & 0.020 & 0.735 & $35.56 \times 11.38$ & $95 \times 220$ & 3.38 & 2.50 \\ \hline
    \end{tabular}
  \end{center}
\end{table*}

\begin{figure}
\epsscale{1.1}
\begin{minipage}[b]{1.0\linewidth}
\plotone{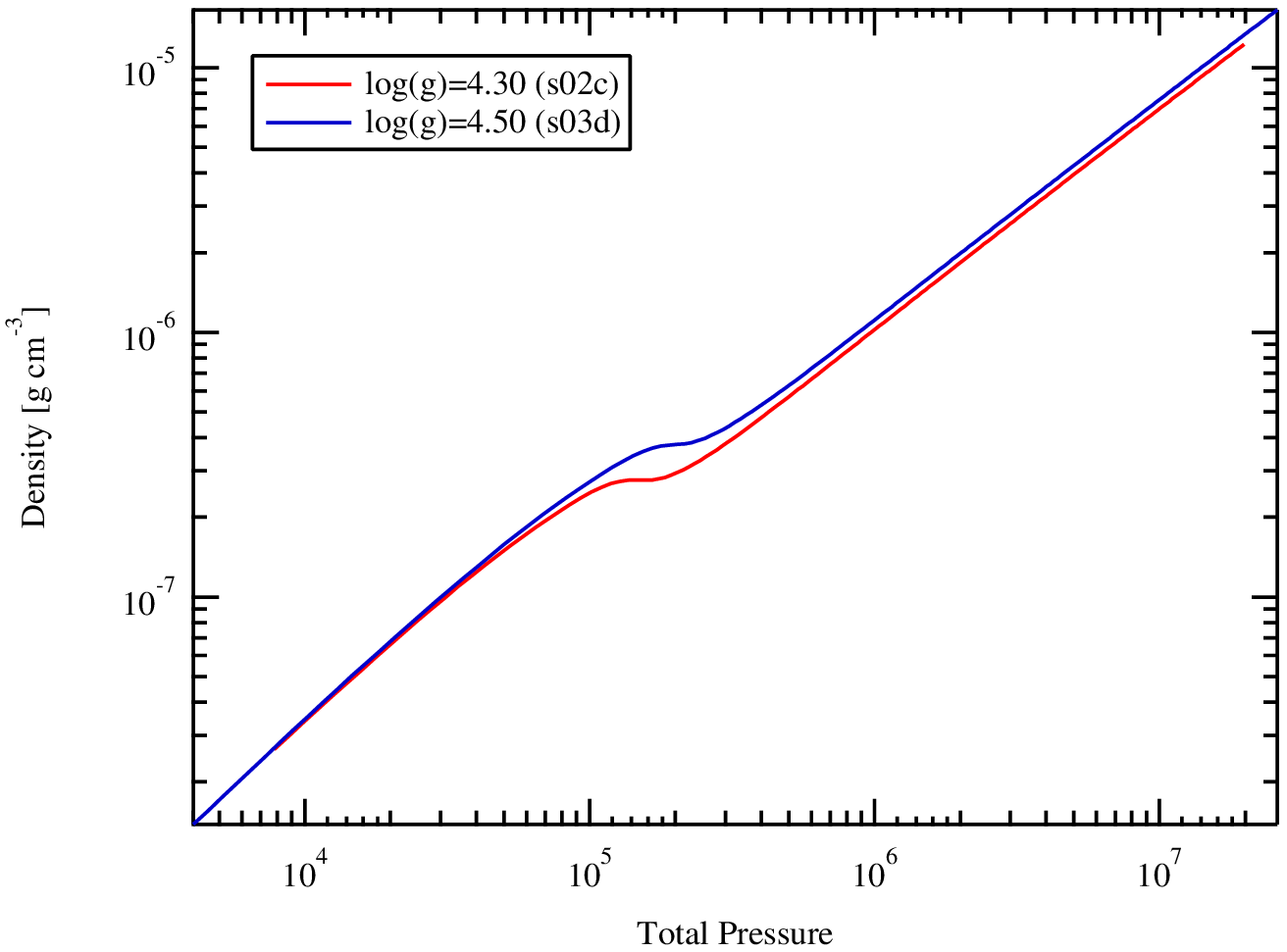}
\epsscale{1.1}
\plotone{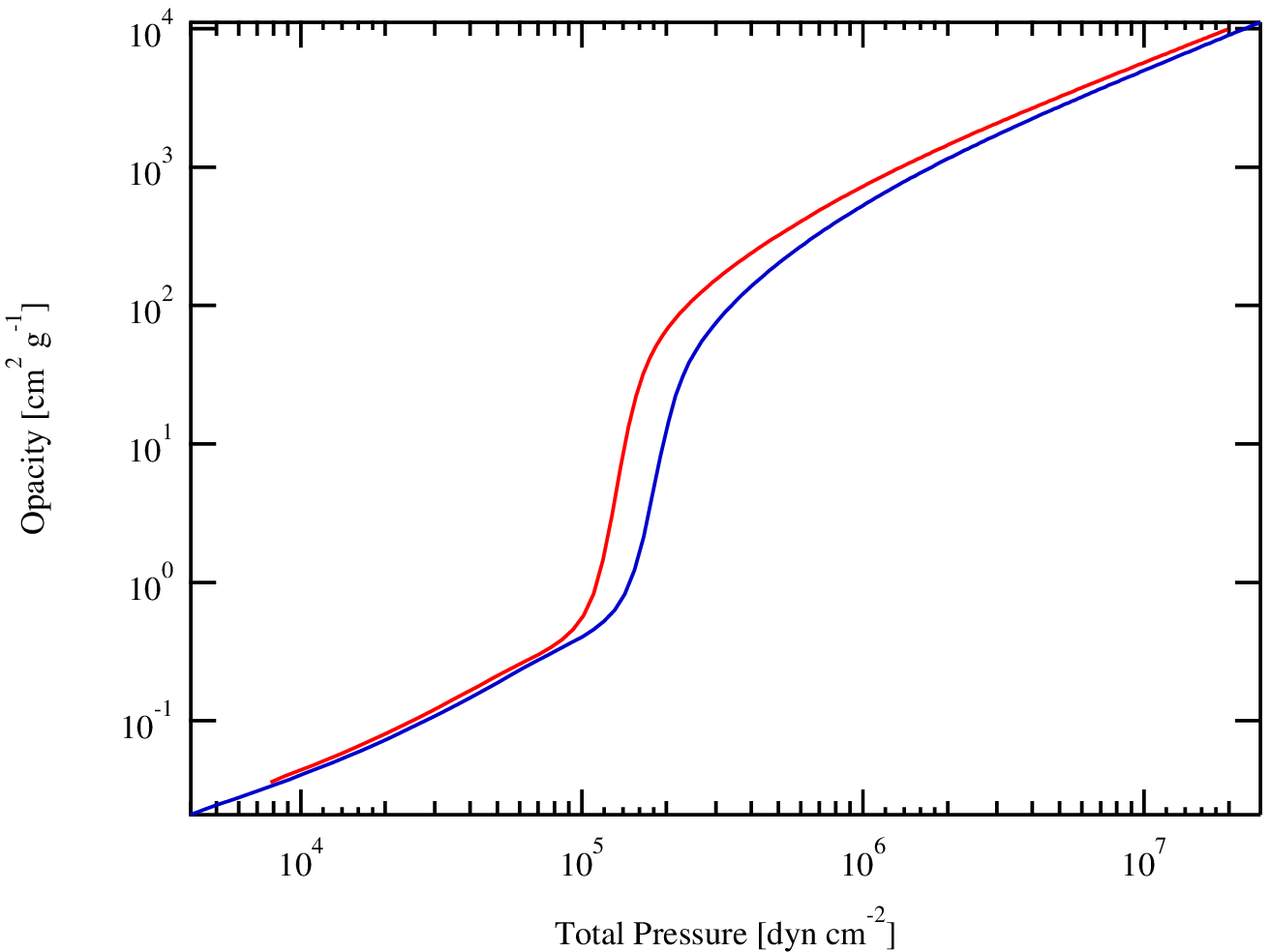}
\epsscale{1.1}
\plotone{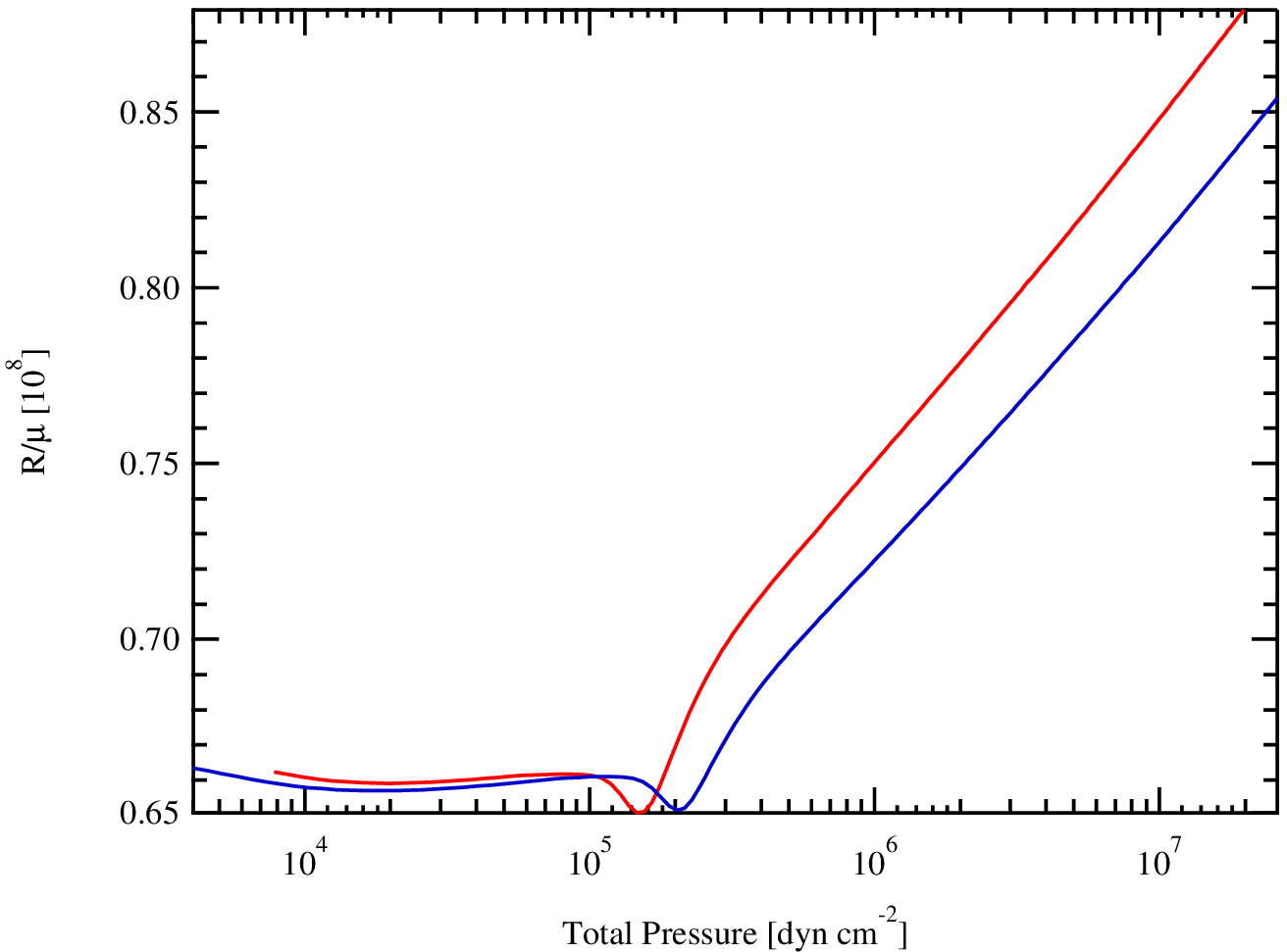}
\end{minipage}
\caption{Mean stratifications of two simulations with different surface gravity.  Their chemical compositions are identical and their effective temperatures are essentially the same, so the differences in structure are attributed to surface gravity.  Comparing this figure to Figs. \ref{fig:opacity},  \ref{fig:mmwt}, and \ref{fig:rhovsp} illustrates that the behaviour of {\logg} and $Y$ are similar below the SAL, but different in the atmosphere.  The differences in the effect of {\logg} and $Y$ are a result of the sensitivity of the mean molecular weight to helium abundance.}
\label{fig:logg3panel}
\end{figure}

\rev{The mean stratification below the SAL behaves as expected, with an increase in {\logg} or $Y$ both yielding a higher density at a given pressure.  Comparing the first panel of Fig. \ref{fig:logg3panel}  to Fig. \ref{fig:rhovsp} confirms that helium and surface gravity qualitatively alter the  structure through the SAL in a similar manner as well.  Differences between the effect of helium and surface gravity are apparent in the atmosphere, however, where for a fixed {\teff} the structures remain similar as {\logg} changes, but are different when helium is changed.  The variation in opacity exhibited in the center panel of Fig. \ref{fig:logg3panel} mirrors the changes to the density and pressure stratification.  Atmospheric opacity remains unchanged between the simulations with varied {\logg}, and below the photosphere the opacity profiles diverge in a similar manner to that seen in Fig. \ref{fig:opacity} from adjusting the helium abundance.}

\rev{The reason that {\logg} and $Y$ affect the sub-photospheric layers in a qualitatively similar manner while behaving differently in the atmospheric layers is because of the mean molecular weight.  Comparing the lower panel of Fig. \ref{fig:logg3panel} to Fig. \ref{fig:mmwt} shows that the mean molecular weight in the atmosphere is sensitive to helium abundance and not surface gravity.}

\rev{In addition to the mean structure, the response of turbulent gas dynamics to {\logg} and $Y$ also show some differences.  In Section \ref{sec:turbulent} (Fig. \ref{fig:pturb_trends}) we did not see a significant sensitivity of the maximum turbulent pressure to helium abundance, although it does depend on metallicity.  In this set of simulations we find that an increase in {\logg} yields a decrease in turbulent pressure, which is qualitatively similar to the results of \citet{2013ApJ...769...18T} and \citet{2013arXiv1302.2621M}.  The degree of convective overshoot (as measured in Section \ref{sec:overshoot}) is also more sensitive to {\logg} than $Y$.}

\section{Discussion and Conclusions}

The signature of helium abundance on stellar convection is smaller than that of metallicity, but is nonetheless significant.  Changes to convective properties are primarily caused by opacity. The sensitivity of convective properties to helium abundance is because of similar reasons as metallicity described in  \citep{2013ApJ...767...78T},  namely, for a given {\teff}, an increase in  opacity in the SAL is accompanied by a corresponding decrease in density so that the constant radiative energy flux is maintained.  Our simulations show systematically larger convective velocities and overshoot with hotter {\teff}, larger $Z$ and smaller $Y$.

While similar, the effect of helium is distinct from that of metallicity because of the way in which it changes the equation of state.  The mean molecular weight is more sensitive to helium abundance, particularly in the optically thin layers above the maximum superadiabaticity.  The signature of helium might manifest itself separately from that of metallicity in the details of the velocity field, such as the velocity distribution presented in Fig. \ref{fig:velocity}.  The key to isolating the effect of helium may be in comparing the hot upflows and cool downdrafts.   Fig. \ref{fig:velocity} shows variation in the `upflow' peak with $Z$, and variation in the extreme edge of the `downflow' peak with $Y$.  Radiation accounts for most of the energy flux from the hot upflowing granules.  Consequently, the structure and dynamics of the granules are more sensitive to radiative cooling rates, which vary more strongly with metallicity than helium.  Conversely, the downflows are less sensitive to radiative cooling rates, but will be affected by the mean molecular weight.  

It is possible that the signature of helium abundance is observable through its effect on the convective overshoot in the atmosphere.  In particular, the curvature of spectral line shapes and the absolute shift of the line cores are sensitive to helium abundance.  Provided that the metallicity \rev{and surface gravity} are well constrained, it might be possible to isolate the effect of helium on the velocity field. \rev{Both surface gravity and helium behave in similar ways below the SAL, but behave differently in the atmosphere.  This is because the helium abundance changes the mean molecular, while surface gravity does not.  In many cases, the signature of helium on the SAL structure and gas dynamics will be smaller than that of metallicity or surface gravity, however, it is distinct in the manner in which it affects mean molecular weight.}

Old metal-poor stellar populations often exhibit signs of $\alpha$-element enrichment.  Using a set of simulations separate from our grid, we have carried out tests to measure the magnitude of the effect of changing the heavy element mixture.  Details of the heavy element mixture alter the opacity because of changes in the number of ionized electrons, which affects H$^-$ opacity.  Our simulations show that the effect is quite small; considerably smaller than the effect of changing $Z$ or $Y$.  While $\alpha$-element enrichment may have a more substantial effect in the deep interior of the star, we do not measure a significant dependence on the structure of the SAL.  Our $\alpha$-enhancement tests also verify that the effect of helium could not be accounted for by adjusting the heavy element mixture.

\acknowledgments

JT is supported by NASA ATFP grant \#NNX09AJ53G to SB, and acknowledges a PGS-D scholarship from the Natural Sciences and Engineering Research Council of Canada. This work was supported in part by the facilities and staff of the Yale University Faculty of Arts and Sciences High Performance Computing Center.


\begin{thebibliography}{88}
\expandafter\ifx\csname natexlab\endcsname\relax\def\natexlab#1{#1}\fi

\bibitem[{Asplund} {et~al.}(1999)]{1999A&A...346L..17A}{{Asplund}, M. and {Nordlund}, {\AA}. and {Trampedach}, R. and {Stein}, R.~F.}, 1999, A\&A, 346, 17

\bibitem[{Asplund} {et~al.}(2000)]{2000A&A...359..669A}{{Asplund}, M., {Ludwig}, H.-G., {Nordlund}, {\AA}. \& {Stein}, R.~F.}, 2000, \aap, 359, 669

\bibitem[{Asplund} {et~al.}(2000)]{2000A&A...359..729A} {{Asplund}, M., {Nordlund}, {\AA}., {Trampedach}, R., {Allende Prieto}, C. \& {Stein}, R.~F.}, 1999, \aap, 359, 729

\bibitem[{Basu} {et~al.}(2004)]{2004ESASP.559..313B} {{Basu}, S., {Mazumdar}, A., {Antia}, H.~M. \& {Demarque}, P. }, 2004, ESASP, 559, 313

\bibitem[{{Beeck} {et~al.}(2012){Beeck}, {Collet}, {Steffen}, {Asplund}, {Cameron}, {Freytag}, {Hayek}, {Ludwig}, \& {Sch{\"u}ssler}}]{2012A&A...539A.121B} {Beeck}, B., {Collet}, R., {Steffen}, M., {et~al.} 2012, \aap, 539, A121

\bibitem[B\"ohm-Vitense(1958)]{1958ZA.....46..108B}B\"ohm-Vitense E., 1958, Z. Astrophys., 46, 108

\bibitem[B\"ohm-Vitense(1989)]{1989isa2.book.....B}B\"ohm-Vitense E., 1989, {Introduction to stellar astrophysics. Vol. 2. Stellar atmospheres.}~Cambridge University Press, Cambridge (UK)

\bibitem[{Bonaca} {et~al.}(2012)]{2012ApJ...755L..12B} {{Bonaca}, A., {Tanner}, J.~D. \& {Basu} { et~al.}}, 2012, \apj, 755, 12

\bibitem[{Chieffi} {et~al.}(1991)]{1991ASPC...13..219C} {{Chieffi}, A., {Straniero}, O. \& {Salaris}, M.}, 1991, ASPCS, 13, 219

\bibitem[{{Chan} \& {Sofia}(1989)}]{1989ApJ...336.1022C} {Chan}, K.~L., \& {Sofia}, S. 1989, \apj, 336, 1022

\bibitem[Canuto \& Dubovikov (1998)]{1998ApJ...493..834C}{{Canuto}, V.~M. and {Dubovikov}, M.}, 1998, \apj, 493, 834

\bibitem[{{Cattaneo} {et~al.}(1991){Cattaneo}, {Brummell}, {Toomre}, {Malagoli}, \& {Hurlburt}}]{1991ApJ...370..282C} {Cattaneo}, F., {Brummell}, N.~H., {Toomre}, J., {Malagoli}, A., \& {Hurlburt},   N.~E. 1991, \apj, 370, 282

\bibitem[{{Collet} {et~al.}(2007){Collet}, {Asplund}, \& {Trampedach}}]{2007A&A...469..687C} {Collet}, R., {Asplund}, M., \& {Trampedach}, R. 2007, \aap, 469, 687

\bibitem[{Collet}(2008)]{2008PhST..133a4004C}{{Collet}, R.}, 2008, PhST, 133

\bibitem[{Collet} {et~al.}(2008)]{2008MmSAI..79..649C} {{Collet}, R., {Asplund}, M. \& {Trampedach}, R.}, 2008, \memsai, 79, 649

\bibitem[{Collet, Magic \& Asplund}(2011)]{2011JPhCS.328a2003C}{{Collet}, R., {Magic}, Z. \& {Asplund}, M.}, 2011, JPhCS, 328, 1

\bibitem[{Chaboyer} {et~al.}(1992)]{1992ApJ...394..515C}{{Chaboyer}, B. and {Sarajedini}, A. \& {Demarque}, P.}, 1992, {\apj}, 394, 515

\bibitem[{{Demarque} {et~al.}(2008){Demarque}, {Guenther}, {Li}, {Mazumdar}, \&   {Straka}}]{2008Ap&SS.316...31D} {Demarque}, P., {Guenther}, D.~B., {Li}, L.~H., {Mazumdar}, A., \& {Straka},   C.~W. 2008, \apss, 316, 31

\bibitem[{Dotter} {et~al}(2008)]{2008ApJS..178...89D} {{Dotter}, A., {Chaboyer}, B., {Jevremovi{\'c}}, D.,  {Kostov}, V., {Baron}, E. \& {Ferguson}, J.~W.}, 2008, 178, 89

\bibitem[{Dravins}(1990)]{1990A&A...228..218D}{{Dravins}, D.}, 1990, {\aap}, 228, 218

\bibitem[{Esch} {et~al.}(2013)]{eschpaper} Esch, L., Bailyn, C., Demarque, P. \& Basu, S., 2013, {\apj}, (submitted).

\bibitem[{{Ferguson} {et~al.}(2005){Ferguson}, {Alexander}, {Allard}, {Barman},   {Bodnarik}, {Hauschildt}, {Heffner-Wong}, \& {Tamanai}}]{2005ApJ...623..585F} {Ferguson}, J.~W., {Alexander}, D.~R., {Allard}, F., {et~al.} 2005, \apj, 623,   585

\bibitem[{{Freytag} {et~al.}(1999){Freytag}, {Ludwig}, \&   {Steffen}}]{1999ASPC..173..225F} {Freytag}, B., {Ludwig}, H.-G., \& {Steffen}, M. 1999, in Astronomical Society   of the Pacific Conference Series, Vol. 173, Stellar Structure: Theory and   Test of Connective Energy Transport, ed. A.~{Gimenez}, E.~F. {Guinan}, \&   B.~{Montesinos}, 225

\bibitem[Gray (1982)]{1982ApJ...255..200G}{{Gray}, D.~F.}, 1982, \apj, 255, 200

\bibitem[Gray (2005)]{graybook} Gray, D.F., 2005, The Observation and Analysis of Stellar Photospheres (3rd ed.); Cambridge Univ. Press

\bibitem[Gray (2009)]{2009ApJ...697.1032G}{{Gray}, D.~F.}, 2009, {\apj}, 697, 1032

\bibitem[Gray (2010a)]{2010ApJ...710.1003G}{{Gray}, D.~F.}, 2010, {\apj}, 710, 1003

\bibitem[Gray (2010b)]{2010ApJ...721..670G}{{Gray}, D.~F.}, 2010, \apj, 721, 670

\bibitem[{{Grevesse} \& {Sauval}(1998)}]{1998SSRv...85..161G} {Grevesse}, N., \& {Sauval}, A.~J. 1998, \ssr, 85, 161

\bibitem[{Grimm-Strele {et~al.}}(2013)]{2013arXiv1305.0743G} {{Grimm-Strele}, H., {Kupka}, F., {L{\"o}w-Baselli}, B.,	{Mundprecht}, E., {Zaussinger}, F. \& {Schiansky}, P.}, 2013, arXiv, 1305.0743

\bibitem[{Harris}(1948)]{1948ApJ...108..112H} {{Harris}, III, D.~L.}, 1948, {\apj}, 108, 112

\bibitem[{Houdek \& Gough}(2007)]{2007MNRAS.375..861H} {{Houdek}, G. \& {Gough}, D.~O.}, 2007, {\mnras}, 375, 861

\bibitem[{Jung} {et ~al.}(2007)]{2007ASPC..362..306J}{{Jung}, Y.~K., {Kim}, Y.-C., {Robinson}, F.~J., {Demarque}, P. \& {Chan}, K.~L.}, 2007, ASPCS, 362, 306

\bibitem[{{Kim} {et~al.}(1995){Kim}, {Fox}, {Sofia}, \&  {Demarque}}]{1995ApJ...442..422K} {Kim}, Y.-C., {Fox}, P.~A., {Sofia}, S., \& {Demarque}, P. 1995, \apj, 442, 422

\bibitem[{{Kim} \& {Chan}(1998)}]{1998ApJ...496L.121K} {Kim}, Y.-C., \& {Chan}, K.~L. 1998, \apjl, 496, L121

\bibitem[Kupka (1999a)]{1999ApJ...526L..45K}{{Kupka}, F.}, 1999, \apjl, 526, 45

\bibitem[Kupka (1999b)]{1999ASPC..173..157K}{{Kupka}, F.}, 1999, ASPCS, 173, 157

\bibitem[{Kupka \& Robinson}(2007)]{2007MNRAS.374..305K} {{Kupka}, F. \& {Robinson}, F.~J.}, 2007, \mnras, 374, 305


\bibitem[{{Kupka} {et~al.}(2009)}]{2009CoAst.160...30K} {{Kupka}, F., {Ballot}, J. \& {Muthsam}, H.~J.}, 2009, CoAst, 160, 30

\bibitem[{{Kupka}(2009)}]{2009MmSAI..80..701K} {Kupka}, F. 2009, \memsai, 80, 701

\bibitem[{{Lee} {et~al.}}(2005)]{2005ApJ...621L..57L} {{Lee}, Y.-W., {Joo}, S.-J., {Han}, S.-I., {Chung}, C., {Ree}, C.~H., {Sohn}, Y.-J., {Kim}, Y.-C., {Yoon}, S.-J., {Yi}, S.~K. \& {Demarque}, P.}, 2005, {\apjl},  621, 57

\bibitem[{{Ludwig} {et~al.}(1998){Ludwig}, {Freytag}, \&   {Steffen}}]{1998IAUS..185..115L} {Ludwig}, H.-G., {Freytag}, B., \& {Steffen}, M. 1998, in IAU Symposium, Vol.   185, New Eyes to See Inside the Sun and Stars, ed. F.-L. {Deubner},   J.~{Christensen-Dalsgaard}, \& D.~{Kurtz}, 115

\bibitem[{{Ludwig} {et~al.}(1999){Ludwig}, {Freytag}, \&   {Steffen}}]{1999A&A...346..111L} {Ludwig}, H.-G., {Freytag}, B., \& {Steffen}, M. 1999, \aap, 346, 111

\bibitem[{{Ludwig} {et~al.}(1995){Ludwig}, {Freytag}, {Steffen}, \&   {Wagenhuber}}]{1995LIACo..32..213L} {Ludwig}, H.-G., {Freytag}, B., {Steffen}, M., \& {Wagenhuber}, J. 1995, in   Li\`ege International Astrophysical Colloquia, Vol.~32, Li\`ege International   Astrophysical Colloquia, 213

\bibitem[{Ludwig \& Steffen}(2008)]{2008psa..conf..133L}{{Ludwig}, H.-G. \& {Steffen}, M.}, 2008, PSA, 133

\bibitem[{Ludwig} {et~al.}(2009)]{2009MmSAI..80..711L}{{Ludwig}, H.-G., {Caffau}, E., {Steffen}, M., {Freytag}, B., {Bonifacio}, P. \& {Ku{\v c}inskas}, A.}, 2009, MmSAI, 80, 711


\bibitem[{Magic} {et~al.}(2013)]{2013arXiv1302.2621M} {{Magic}, Z., {Collet}, R., {Asplund}, M., {Trampedach}, R.,  {Hayek}, W., {Chiavassa}, A., {Stein}, R.~F. \& {Nordlund}, {\AA}. }, 2013, \aap, 557, A26

\bibitem[{{Nordlund}(1985{\natexlab{b}})}]{1985SoPh..100..209N} {{Nordlund}, A.}, 1985, \solphys, 100, 209

\bibitem[{{Nordlund}(1982)}]{1982A&A...107....1N} {Nordlund}, A. 1982, \aap, 107, 1

\bibitem[Nordlund \& Dravins (1990)]{1990A&A...228..155N}{{Nordlund}, A. and {Dravins}, D.}, 1990, \aap, 228, 155

\bibitem[{Prakapavicius} {et~al.}(2013)]{2013arXiv1303.2016P} {{Prakapavicius}, D., {Steffen}, M., {Kucinskas}, A., {Ludwig}, H.-G., {Freytag}, B., {Caffau}, E. \& {Cayrel}, R.}, 2013, arXiv, 1303.2016

\bibitem[{{Robinson} {et~al.}(2003){Robinson}, {Demarque}, {Li}, {Sofia},   {Kim}, {Chan}, \& {Guenther}}]{2003MNRAS.340..923R} {Robinson}, F.~J., {Demarque}, P., {Li}, L.~H., {et~al.} 2003, \mnras, 340, 923

\bibitem[{{Robinson} {et~al.}(2004){Robinson}, {Demarque}, {Li}, {Sofia},   {Kim}, {Chan}, \& {Guenther}}]{2004MNRAS.347.1208R} ---. 2004, \mnras, 347, 1208

\bibitem[{{Robinson} {et~al.}(2005){Robinson}, F.~J., {Demarque}, P., {Guenther}, D.~B., {Kim}, Y.-C. \& {Chan}, K.~L.}]{2005MNRAS.362.1031R}, 2005, MNRAS, 362, 1031

\bibitem[{{Rogers} \& {Nayfonov}(2002)}]{2002ApJ...576.1064R} {Rogers}, F.~J., \& {Nayfonov}, A. 2002, \apj, 576, 1064

\bibitem[{Rosenthal} {et~al.}(1999)]{1999A&A...351..689R}{{Rosenthal}, C.~S., {Christensen-Dalsgaard}, J., {Nordlund}, {\AA}.,	{Stein}, R.~F. \& {Trampedach}, R.}, 1999, A\&A, 351, 689

\bibitem[{Salaris} {et~al.}(1993)]{1993ApJ...414..580S} {{Salaris}, M., {Chieffi}, A. \& {Straniero}, O.}, 1993, \apj, 414, 580

\bibitem[{{Stein} \& {Nordlund}(1989)}]{1989ApJ...342L..95S} {Stein}, R.~F., \& {Nordlund}, A. 1989, \apjl, 342, L95

\bibitem[{{Stein} \& {Nordlund}(1998)}]{1998ApJ...499..914S} {{Stein}, R.~F. and {Nordlund}, A.}, 1998, \apj, 499, 914

\bibitem[{Stein} {et~al.}(1999)]{1999AAS...194.2104S}{{Stein}, R.~F. and {Bercik}, D. and {Georgobiani}, D. and {Nordlund}, {\AA}.}, 1999, AAS, 194, 2014

\bibitem[{{Stein} \& {Nordlund}(2000)}]{2000SoPh..192...91S} {Stein}, R.~F., \& {Nordlund}, {\AA}. 2000, \solphys, 192, 91

\bibitem[{Sweigart}(1997)]{1997ApJ...474L..23S}{{Sweigart}, A.~V.}, 1997, {\apjl}, 474, 23

\bibitem[{Sweigart \& Mengel}(1979)]{1979ApJ...229..624S}{{Sweigart}, A.~V. \& {Mengel}, J.~G.}, 1979, {\apj}, 229, 624

\bibitem[{Tanner} {et~al.}(2012)]{2012ApJ...759..120T}{{Tanner}, J.~D., {Basu}, S. \& {Demarque}, P.}, 2012, \apj, 759, 120

\bibitem[{Tanner} {et~al.}(2013)]{2013ApJ...767...78T}{{Tanner}, J.~D., {Basu}, S. \& {Demarque}, P.}, 2013, \apj, 767, 78

\bibitem[Trampedach (2010)]{2010Ap&SS.328..213T}{{Trampedach}, R.}, 2010, \apss, 328, 213

\bibitem[Trampedach {et al.}(2013)]{2013ApJ...769...18T} {{Trampedach}, R., {Asplund}, M., {Collet}, R., {Nordlund}, {\AA}. \ {Stein}, R.~F.}, 2013, \apj, 769, 18

\bibitem[{{Unno} \& {Spiegel}(1966)}]{1966PASJ...18...85U} {Unno}, W., \& {Spiegel}, E.~A. 1966, \pasj, 18, 85

\bibitem[{Zahn}(1992)]{1992A&A...265..115Z} {{Zahn}, J.-P.}, 1992, {\aap}, 265, 115

\end{thebibliography}
\end{document}